\documentclass[manuscript,screen,review=false]{acmart}

\usepackage{multicol,multirow}
\usepackage{color}
\usepackage{xcolor}
\usepackage{colortbl}
\usepackage{float}
\usepackage{makecell}
\usepackage{cellspace}
\usepackage{booktabs}
\definecolor{mygray}{gray}{0.6}
\definecolor{myblue}{rgb}{0.8,0.85,1}
\usepackage{array}
\newcolumntype{L}[1]{>{\raggedright\let\newline\\\arraybackslash\hspace{0pt}}m{#1}}
\newcolumntype{C}[1]{>{\centering\let\newline\\\arraybackslash\hspace{0pt}}m{#1}}
\newcolumntype{R}[1]{>{\raggedleft\let\newline\\\arraybackslash\hspace{0pt}}m{#1}}
\newcommand{\noun}[1]{\textsc{#1}}

\newtheorem{definition}{Definition}

\newtheorem{target}{Target}
\newtheorem{challenge}{Challenge}

\AtBeginDocument{%
  \providecommand\BibTeX{{%
    \normalfont B\kern-0.5em{\scshape i\kern-0.25em b}\kern-0.8em\TeX}}}

\setcopyright{acmcopyright}
\copyrightyear{2023}
\acmYear{2023}
\acmDOI{XXXXXXX.XXXXXXX}

\begin{document}

\title{A Survey on Federated Unlearning: Challenges, Methods, and Future Directions}

\author{Ziyao Liu}
\email{liuziyao@ntu.edu.sg}
\affiliation{%
  \institution{Digital Trust Centre, Nanyang Technological University}
  \country{Singapore}
}

\author{Yu Jiang}
\email{yu012@e.ntu.edu.sg}
\author{Jiyuan Shen}
\email{jiyuan001@e.ntu.edu.sg}
\author{Minyi Peng}
\email{minyi002@e.ntu.edu.sg}
\affiliation{%
  \institution{College of Computing and Data Science, Nanyang Technological University}
  \country{Singapore}}

\author{Kwok-Yan Lam}
\email{kwokyan.lam@ntu.edu.sg}
\affiliation{%
  \institution{College of Computing and Data Science and Digital Trust Centre, Nanyang Technological University}
  \country{Singapore}
  }

\author{Xingliang Yuan}
\email{xingliang.yuan@unimelb.edu.au}
\affiliation{%
  \institution{School of Computing and Information Systems, The University of Melbourne}
  \country{Australia}
  }

\author{Xiaoning Liu}
\email{xiaoning.liu@rmit.edu.au}
\affiliation{%
  \institution{School of Computing Technologies, RMIT University}
  \country{Australia}
  }


\begin{abstract}
In recent years, the notion of ``the right to be forgotten" (RTBF) has become a crucial aspect of data privacy for digital trust and AI safety, requiring the provision of mechanisms that support the removal of personal data of individuals upon their requests. Consequently, machine unlearning (MU) has gained considerable attention which allows an ML model to selectively eliminate identifiable information. Evolving from MU, federated unlearning (FU) has emerged to confront the challenge of data erasure within federated learning (FL) settings, which empowers the FL model to unlearn an FL client or identifiable information pertaining to the client. Nevertheless, the distinctive attributes of federated learning introduce specific challenges for FU techniques. These challenges necessitate a tailored design when developing FU algorithms. While various concepts and numerous federated unlearning schemes exist in this field, the unified workflow and tailored design of FU are not yet well understood. Therefore, this comprehensive survey delves into the techniques and methodologies in FU providing an overview of fundamental concepts and principles, evaluating existing federated unlearning algorithms, and reviewing optimizations tailored to federated learning. Additionally, it discusses practical applications and assesses their limitations. Finally, it outlines promising directions for future research.
\end{abstract}



\received{20 November 2023}
\received[revised]{20 November 2023}
\received[accepted]{20 November 2023}

\maketitle

\section{Introduction}
\label{sec:introduction}
\begin{figure}[htbp!]
\centering
		\centering
		\includegraphics[width=0.6\linewidth]{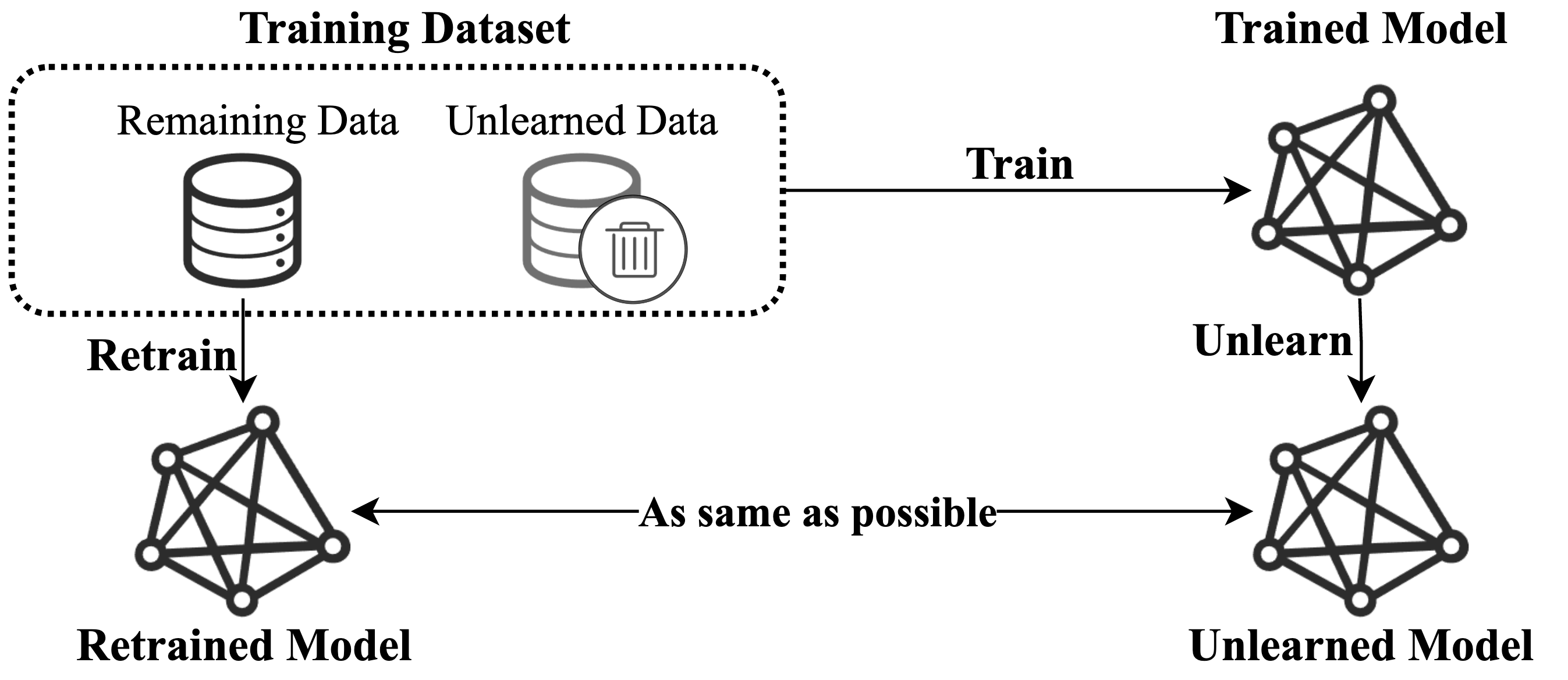}
		\caption{Machine unlearning. Naive retraining, discarding the trained model and starting training from scratch with remaining data after unlearned data removal, is computationally intensive. Conversely, machine unlearning, which resumes training from the trained model through an unlearning process, is much more cost-effective. The objective of MU is to ensure that the unlearned model achieves a performance level on par with that of the retrained model.}
		\label{fig:MU-scheme}
\end{figure}

With increasing concerns for personal data privacy protection, governments and legislators around the world have enacted rigorous data privacy regulations, such as GDPR \cite{regulation2018general}, APPI \cite{iwase2019overview} and CCPA \cite{goldman2020introduction}. Typically, as digital service providers capture personal data from their users or data owners for service development, such regulations require them to grant users the right to be forgotten (RTBF), with the provision of mechanisms that allow them to request the removal of their personal data from digital records. Consequently, given the extensive adoption of data-intensive machine learning (ML) algorithms, RTBF enables users to purge their data, including the influence of these data, from both the training dataset and the trained ML model. This is where machine unlearning (MU) \cite{xu2023machine,bourtoule2021machine,huynh2024fast,hu2024learn,chen2021machine,han2024towards,hu2024duty,nguyen2022survey,liu2023breaking,liu2024towards,hu2023eraser,wang2023machine} steps in as a critical facilitator of this process, ensuring that personal data is effectively and responsibly removed, further strengthening data privacy and ethical data handling. As depicted in Figure \ref{fig:MU-scheme}, the primary objective of unlearning is to remove the impact of specific data points from a trained model, while preserving the overall performance of the model.

\begin{figure}
\centering
	\centering
		\includegraphics[width=0.8\linewidth]{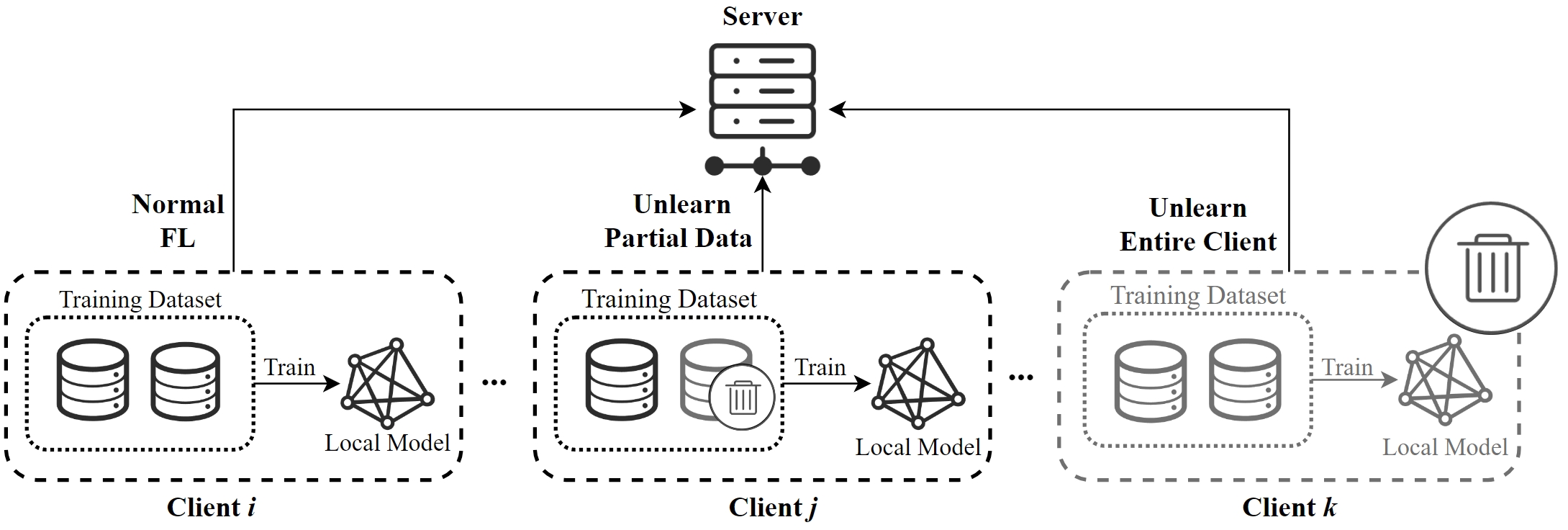}
		\caption{Federated unlearning. In contrast to machine unlearning algorithms, which are typically executed by a single entity, FU systems involve multiple entities, including the unlearned client, remaining clients, and the central server, any of whom can act as the unlearner, responsible for executing the unlearning algorithm. Furthermore, the unlearning target may encompass either an entire client or specific partial data from a target client.}
		\label{fig:FU-scheme}
\end{figure}

Building upon the core principles of MU and the concept of RTBF, federated unlearning (FU) \cite{cao2023fedrecover,liu2022right,ding2023strategic,tao2024communication,wang2024server,jiang2024towards,guo2023fast,romandini2024federated,qiu2023fedcio,wang2023bfu,yuan2024towards} has emerged to confront the challenge of data erasure within the domain of federated learning (FL) settings \cite{kairouz2021advances,mcmahan2017communication,yang2019federated,zhang2023agrevader,wu2024cardinality}. In a typical FL system, multiple clients locally train their machine learning models, which are subsequently aggregated to construct a global model. Then the server distributes the updated global model to all clients for training in the subsequent FL round. These sequential steps continue to recur until the global model reaches convergence. (see Section \ref{sec:federated_learning} for more details). As a result, the objective of FU is to enable the FL model to remove the impact of an FL client or identifiable information associated with a client's partial data, while maintaining the privacy guarantees of the decentralized learning process, as illustrated in Figure \ref{fig:FU-scheme}. A formal definition of FU is provided in Section \ref{sec:fu_definition}.

However, in contrast to traditional machine unlearning, the unique characteristics of federated learning introduce new targets and challenges. (see Section \ref{sec:targets_and_challenges_of_federated_unlearning} for more details). Therefore, this survey delves into techniques, methodologies, and recent advancements in federated unlearning. We provide an overview of fundamental concepts and principles in FU design, evaluate existing FU algorithms, present a taxonomy, and review optimizations of FU tailored to federated learning settings.

\textbf{Comparison with related surveys.}
Currently, there are some works that have been conducted to summarize machine unlearning \cite{nguyen2022survey,xuh2023machine, xu2023machine,shaik2023exploring,qu2023learn,liu2024threats,wang2024machine}. However, few existing surveys perceive the construction of federated unlearning. In \cite{wu2023knowledge}, the concept of knowledge editing throughout the entire lifecycle of federated learning is explored. This survey categorizes relevant works based on the principles of exact learning and approximate learning. These categories, as described in previous machine unlearning taxonomies like those in \cite{xuh2023machine, xu2023machine}, are not specifically designed for a federated setting and thus may not fully capture the unique characteristics inherent in FU designs. The survey conducted in \cite{wang2023federated} focuses on an analysis of only privacy and security threats within FU systems, with extensive discussions on potential attacks and defensive measures. It pays particular attention to the issue of privacy leakage stemming from distinctions between the trained model and the unlearned model, specifically examining their vulnerability to membership inference attacks. \cite{yang2023survey} provides a brief survey on federated unlearning, focusing on the level of data erasure, similar to the "unlearn-what" aspect discussed in our work. The works most closely related to ours are \cite{romandini2024federated} and \cite{jeong2024sok}, which provide comprehensive surveys on existing federated unlearning literature. However, \cite{romandini2024federated} lacks an investigation into FL-tailored optimization and the limitations of existing approaches, while \cite{jeong2024sok} lacks a formal definition of federated unlearning. Additionally, both \cite{romandini2024federated} and \cite{jeong2024sok} do not describe the unlearning workflow, which is important for readers to understand how unlearning integrates with Machine Learning as a Service (MLaaS). Furthermore, they do not specifically focus on security and privacy issues in federated unlearning systems.

While various concepts and numerous federated unlearning schemes exist in this field, the design and implementation of FU are still not fully explored. Furthermore, the methodology and principles for extending machine unlearning approaches to federated unlearning remain relatively unclear. The unified workflow of FU, particularly regarding security and privacy issues, is not yet well understood. This lack of comprehensive resources serves as the primary motivation for our effort in delivering this survey, which offers a deep and thorough insight into current FU research. A detailed comparison of related FU surveys is summarized in Table \ref{tab:survey_comp}.

\begin{table}[htbp!]
\centering
\begin{tabular}{|c|cc|cccc|cccc|cccc|}
\hline
\multicolumn{1}{|c|}{Ref.}&\multicolumn{2}{|c|}{Def.}&\multicolumn{4}{|c|}{Taxonomy}&\multicolumn{4}{|c|}{Review}&\multicolumn{4}{|c|}{Insight}\\
\hline
& \rotatebox{90}{Target Formalization} & \rotatebox{90}{Summary of Challenges} & \rotatebox{90}{Unlearning Workflow} & \rotatebox{90}{Who-unlearn} & \rotatebox{90}{Unlearn-what} & \rotatebox{90}{Who-verify} & \rotatebox{90}{Comprehensive Review} & \rotatebox{90}{Principle Analysis} & \rotatebox{90}{Security \& Privacy} & \rotatebox{90}{Proof of Unlearning} & \rotatebox{90}{FL-tailored optimization} & \rotatebox{90}{Limitation} & \rotatebox{90}{Experimental Evaluation} & \rotatebox{90}{Future Directions}  \\ \hline
Wang et al. \cite{wang2023federated}&-&\checkmark&-&\checkmark&-&-&-&-&\checkmark&-&-&-&-&-\\
Wu et al. \cite{wu2023knowledge}&-&\checkmark&\checkmark&-&-&-&-&\checkmark&-&\checkmark&\checkmark&-&-&\checkmark\\
Yang and Zhao \cite{yang2023survey}&\checkmark&\checkmark&-&-&\checkmark&-&-&-&-&\checkmark&\checkmark&-&\checkmark&\checkmark\\
Nicol{\`o} et al. \cite{romandini2024federated}&\checkmark&\checkmark&-&\checkmark&\checkmark&-&\checkmark&\checkmark&-&\checkmark&-&-&\checkmark&\checkmark\\
Jeong et al. \cite{jeong2024sok}&-&\checkmark&-&\checkmark&\checkmark&-&\checkmark&\checkmark&-&\checkmark&\checkmark&\checkmark&-&\checkmark\\
\textbf{Ours}&\checkmark&\checkmark&\checkmark&\checkmark&\checkmark&\checkmark&\checkmark&\checkmark&\checkmark&\checkmark&\checkmark&\checkmark&-&\checkmark\\
\hline
\end{tabular}
\caption{Comparison of related FU surveys.}
\label{tab:survey_comp}
\end{table}

\textbf{Summary of contributions.} The main contributions of this survey are listed as follows.
\begin{enumerate}
    \item We present a unified federated unlearning workflow, on the basis of which we offer a novel taxonomy of existing FU techniques.
    \item Utilizing the proposed taxonomy and considering factors including (i) who-unlearn and (ii) unlearn-what, we conduct a comprehensive summary of existing federated unlearning methods, and highlight their distinctions, advantages, and constraints.
    \item We conduct a comprehensive examination of optimizations of FU techniques specifically tailored for federated learning, along with an assessment of their limitations.
    \item We delve deeply into critical discussions concerning the existing challenges in federated unlearning, and identify promising directions for future research.
\end{enumerate}

\begin{figure}[htbp!]
    \centering
    \includegraphics[width=0.9\linewidth]{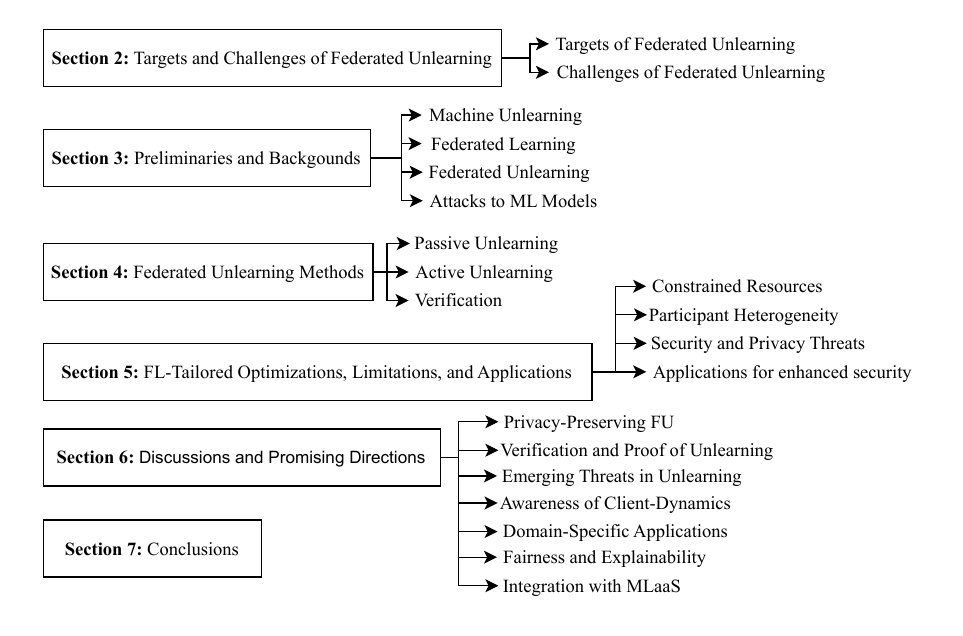}
    \caption{Illustrative organization of the paper.}
    \label{fig:organization}
\end{figure}

\textbf{Organization of the paper.} The rest of this paper is organized as follows. Section \ref{sec:targets_and_challenges_of_federated_unlearning} summarizes the targets, challenges, and characteristics of federated unlearning, and discusses their alignment. Section \ref{sec:preliminaries_and_backgrounds} describes the principles employed to achieve machine unlearning and provides an overview of the fundamentals of federated learning and unlearning. Section \ref{sec:federated_unlearning_methods} presents different constructions of existing FU algorithms, followed by reviews of various optimizations tailored to federated learning and a critical examination of their limitations in Section \ref{sec:fl_specific_considerations}. Section \ref{sec:discussions_and_promising_directions} offers discussions and outlines future research directions. Finally, Section \ref{sec:conclusions} summarizes and concludes the paper. An illustrative organization of the paper is provided in Figure \ref{fig:organization}.

\section{Targets and Challenges of Federated Unlearning}
\label{sec:targets_and_challenges_of_federated_unlearning}

In this section, we will explore the targets of federated unlearning and the associated challenges compared to traditional machine unlearning. The insights gained will serve as a guideline for the taxonomy presented in Section \ref{sec:federated_unlearning_methods}.

\subsection{Targets of Federated Unlearning}

We now specify the targets of the unlearning process within an FL setting, for which the formal definitions are provided in Section \ref{sec:fu_definition}.
\begin{target}\textbf{(Model consistency)}
    The unlearned model must exhibit performance akin to a retrained model, ensuring the unlearning process neither diminishes its accuracy nor reliability. Achieving this consistency is crucial, as it demonstrates the effectiveness of the unlearning algorithm in removing specific data while maintaining the model's overall quality.
\end{target}
\begin{target}\textbf{(Unlearning efficiency)}
    As retraining in an FL system involves starting the training process from scratch, which is often inefficient, the target related to unlearning efficiency is to ensure that the cost of unlearning is significantly lower than that of obtaining a retrained model. These costs encompass various factors such as runtime, the number of participating clients, and communication overhead.
\end{target}
\begin{target}\textbf{(Privacy preservation)}
    FL is designed to offer privacy assurances by allowing access only to the locally trained model rather than the local data. Therefore, unlearning in a federated context must also ensure the preservation of clients' local data privacy. This approach ensures that while unlearning processes are implemented, the fundamental privacy guarantees of FL are maintained, safeguarding the privacy of clients' local data.
\end{target}
\begin{target}\textbf{(Certified removal)}
    The capability to verify the removal of either an entire FL client or partial data from a target client is essential. This process of certified removal should align with the unlearning request made by an FL participant. In other words, if the unlearning request is raised by an FL client, this client must be allowed to verify if its data has been unlearned and its impact on the FL model has been removed. Similarly, if the server raises the unlearning request, the server must also be able to monitor and verify the unlearning process. This verification process must be robust and reliable, ensuring that the removal adheres strictly to the specified unlearning request, thereby maintaining the trustworthiness of the federated unlearning system.
\end{target}

\subsection{Challenges of Federated Unlearning}

In contrast to traditional machine unlearning, the unique characteristics of federated learning introduce certain challenges to the unlearning technique, as outlined below.

\begin{challenge}
    \textbf{(Knowledge Permeation)}
    When a client's data needs to be unlearned, its information has already spread throughout all participants in the FL system. This occurs because, during each FL round, the server aggregates the gradients from all clients and updates the global model. This updated model is then distributed to all clients, on which all clients conduct the subsequent round of FL training. As a result, the knowledge from the targeted client for unlearning permeates through to the other clients via the FL training process. Consequently, knowledge permeation complicates the achievement of model consistency (\textbf{Target 1}) in FU, compared to data-centralized MU schemes. Furthermore, implementing unlearning in a federated setting requires the involvement of all impacted clients, which significantly increases associated costs and impacts the target of unlearning efficiency (\textbf{Target 2}).
\end{challenge}
\begin{challenge}
    \textbf{(Data Isolation)}
    Since every client individually maintains its dataset and conducts local model training, which is a key advantage of FL in terms of privacy preservation, only gradients or global models are publicly shared in an FL system. This aspect might hinder adapting existing MU algorithms, which rely on direct data access to be unlearned, within the FL context, aligning with the privacy preservation target (\textbf{Target 3}). Moreover, the absence of direct access to the unlearned data poses challenges in creating efficient FU algorithms, leading to concerns about unlearning efficiency (\textbf{Target 2}), compared to more efficient MU algorithms that directly utilize the unlearned data.
\end{challenge}
\begin{challenge}
    \textbf{(Who-Unlearn)}
    Different from machine unlearning algorithms, which are typically executed by a single client, FU systems involve multiple participants, including (i) the unlearned client or target client\footnote{We use unlearned client and target client interchangeably.}, (ii) remaining clients, and (iii) the central server, any of whom can act as the unlearner, responsible for executing the unlearning algorithm. Therefore, the FU algorithm selected by the unlearner depends on the degree of access to information about the data to be unlearned, consistent with the target of privacy preservation (\textbf{Target 3}). For example, when unlearning partial data of a target client, the target client possesses direct access to both the unlearned and remaining data, while the server's access is limited to historical data in the form of global models and gradients. Furthermore, when a client initiates a request for unlearning, it has the option to either participate in the unlearning process or simply exit the system. In cases where the target client chooses to leave, the unlearning process can be executed either on the server, the remaining clients, or both. Additionally, the entity responsible for unlearning also influences the need to verify that the unlearning process has been executed in accordance with the unlearner's request, contributing to the target of certified removal (\textbf{Target 4}).
\end{challenge}
\begin{challenge}
    \textbf{(Unlearn-What)}
    In an FU system, the initiation of an unlearning request can stem from either the unlearned client or the server for different purposes. Concurrently, it's essential to consider that the unlearning target can be either (i) an entire target client or (ii) specific partial data from a target client. Considering the ``unlearn-what" aspect, the unlearning principles of FU algorithms differ significantly. For example, when unlearning an entire target client, methods like local retraining, fine-tuning, and multi-task unlearning are no longer applicable (see Section \ref{sec:unlearning_principles} for more details). This is because these FU algorithms rely on direct access to the unlearned data and the remaining data of the target client, which becomes inaccessible when the entire client needs to be removed, in accordance with the target of privacy preservation removal (\textbf{Target 3}). The variation in designing FU algorithms for different unlearning targets also impacts the performance in achieving model consistency (\textbf{Target 1}), unlearning efficiency (\textbf{Target 2}), and certified removal (\textbf{Target 4}).
\end{challenge}
\begin{challenge}
    \textbf{(Who-Verify)}
    As an FU system involves multiple participants, unlearning requests may be raised by FL clients or the FL server. For a more compatible scenario with the RTBF regulations, where unlearning services are provided by MLaaS infrastructures, clients must be allowed to verify if their data has been unlearned and its impact on the FL model has been removed. Similarly, robust and provable proof of unlearning should be conducted on the server side if the request is raised by the server. However, implementing these verification processes presents significant challenges, including ensuring the efficiency (\textbf{Target 2}) and reliability (\textbf{Target 4}) of verification methods, maintaining system performance, and addressing potential security vulnerabilities. Only when data removal adheres strictly to the specified unlearning request can the trustworthiness of the federated unlearning system be maintained.
\end{challenge}

In addition to the primary distinctions and challenges highlighted above in FU in comparison to MU, several other factors may impede the effectiveness of federated unlearning. These factors arise from the unique characteristics of FL systems and are outlined as follows. Table \ref{tab:target_alignment} summarizes the alignment between these targets, challenges, and characteristics.

\begin{enumerate}
    \item \textbf{Constrained Resources}: In federated learning, devices or nodes that engage in the process often contend with constraints on their computing power, communication capabilities (such as limited network bandwidth), and storage capacities (like constrained memory). These limitations can affect their capacity to execute intricate model training tasks, facilitate efficient sharing and reception of updates, as well as manage the storage and processing of large machine learning models, datasets, or supplementary information. Consequently, resource-intensive MU algorithms may no longer be practical or scalable within the context of federated learning.
    \item \textbf{Participant Heterogeneity}: In FL systems, clients exhibit heterogeneity in various aspects, including their training capabilities related to factors such as data structure and distributions, e.g., vertical partitioned features and non-identically distributed data (Non-IID) data. This diversity necessitates the development of heterogeneity-aware FU approaches.
    \item \textbf{Client Dynamics}: In each FL round, clients are randomly chosen to participate in the model aggregation process. Besides, there may be a large number of dropped clients and newly-joined clients. The unlearner faces significant challenges in recalling past clients for unlearning operations, let alone retraining the model from scratch. These dynamic client behaviors can exert an influence on the effectiveness of machine unlearning algorithms, which were initially tailored for scenarios involving a single client in MU settings.
    \item \textbf{Security and Privacy Threats}: In FU settings, malicious attacks and information leakage are more intricate compared to a single-client MU scenario. Threat models become increasingly complex, taking into account factors like adversaries, their capabilities, and the potential for collusion.
\end{enumerate}

\begin{table}[htbp!]
\centering
\begin{tabular}{@{}l|lllll|llll@{}}
\toprule
\multicolumn{1}{c|}{Target} & \multicolumn{5}{c|}{Challenge} & \multicolumn{4}{c}{Characteristic}\\
\hline
\multicolumn{1}{c|}{} & \rotatebox{90}{Knowledge Permeation} & \rotatebox{90}{Data Isolation} & \rotatebox{90}{Who-Unlearn} & \rotatebox{90}{Unlearn-What} & \rotatebox{90}{Who-Verify} & \rotatebox{90}{Constrained Resources} & \rotatebox{90}{Participant Heterogeneity} & \rotatebox{90}{Client Dynamics} & \rotatebox{90}{Security \& Privacy} \\ \midrule
Model Consistency & \checkmark &  &  & \checkmark &  &  &  & \checkmark &  \\
Unlearning Efficiency & \checkmark & \checkmark &  & \checkmark & \checkmark & \checkmark & \checkmark &  &  \\
Privacy Preservation &  & \checkmark & \checkmark & \checkmark &  &  &  & \checkmark & \checkmark \\
Certified Removal &  &  & \checkmark & \checkmark & \checkmark &  &  & \checkmark &  \\ \bottomrule
\end{tabular}
\caption{Alignment between targets, challenges and characteristics of federated unlearning.}
\label{tab:target_alignment}
\end{table}

\section{Preliminaries and Backgrounds}
\label{sec:preliminaries_and_backgrounds}

In this section, we will first provide an overview of machine unlearning and summarize the principles of unlearning algorithms and metrics for the verification of unlearning. Then, we will provide an overview and formalization of federated learning and federated unlearning. Since attacks on machine learning models can be used for the verification of unlearning, an additional subsection is included to introduce attacks on ML models for completeness.

\subsection{Machine Unlearning}
\label{sec:machine_unlearning}

In the MU system, the training dataset $D$ consists of two components: $D_u$, representing the data samples to be forgotten, and $D_r$, representing the remaining data samples, where $D_r = D \backslash D_u$. We then consider $\mathcal{M}(D)$ as the final model trained on dataset $D$. 

\subsubsection{Unlearning principles}
\label{sec:unlearning_principles}
Existing MU research papers predominantly rely on the following unlearning principles to make the distribution of the model $\mathcal{M}(D)$ identical to the distribution of the model $\mathcal{M}(D_r)$ \cite{bourtoule2021machine}.

\textbf{Retraining.} is a process training from a model free from the influence of data from $D_u$ on the dataset $D_r$, essentially starting from scratch.
In this method, a newly trained model $\mathcal{M}(D_r)$ does not have any information about $D_u$. However, this process is both time-consuming and resource-intensive because it discards the model $\mathcal{M}(D)$ on $D$ containing the contribution of $D_r$ dataset.

\textbf{Fine-tuning.} uses the remaining dataset $D_r$ to optimize the model $\mathcal{M}(D)$ and reduce the impact of data from $D_u$. However, this process involves multiple iterations, leading to increased computational and communication costs.

\textbf{Gradient ascent.} represents a reverse learning process. In machine learning, the model $\mathcal{M}(D)$ is trained by minimizing the loss using gradient descent. Conversely, the unlearning process involves the application of gradient ascent to maximize the loss. However, this method can easily lead to catastrophic forgetting. As a result, many studies introduce constraints to preserve memory.

\textbf{Multi-task unlearning} seeks to not only eliminate the influence of $D_u$ but also to reinforce the acquisition of knowledge from the remaining data $D_r$. In the course of these endeavors, most studies aim to strike a balance between the erasure effect and the retention effect.

\textbf{Model scrubbing.} applies a ``scrubbing" transformation $\mathscr{H}$ to the model $\mathcal{M}(D)$ to ensure that the unlearned model closely approximates the perfectly retrained model with only $D_r$, as expressed by $\mathscr{H}(\mathcal{M}(D)) \approx \mathcal{M}(D_r)$ \cite{ginart2019making}. When defining the scrubbing method $\mathscr{H}$, most approaches rely on a quadratic approximation of the loss function. Specifically, for model parameters $\theta$ and $\phi$, the gradient of the loss function of a given data point $D_x$ satisfies
$$\nabla f_{D_x}(\phi) = \nabla f_{D_x}(\theta) + \mathcal{H}_{D_x}(\theta)(\phi-\theta),$$ 
where $\mathcal{H}_{D_x}(\theta)$ is positive semi-define. The scrubbed model becomes the new optimum by setting $\nabla f_{D_r}(\mathscr{H}_{D_r}(\theta)) = 0$, yielding the equation:
$$\mathscr{H}_{D_r}(\theta) = \theta - \mathcal{H}_{D_r}^{-1} (\theta) \nabla f_{D_r}(\theta).$$
$\mathscr{H}$ can perform a Newton step and can be derived under various theoretical assumptions \cite{fraboni2022sequential} \cite{golatkar2020forgetting}. However, the challenge of this method lies in computing the Hessian matrix, which is infeasible for high-dimensional models. Therefore, some approaches aim to compute an approximation of the Hessian.

\textbf{Synthetic data.} is a method that replaces certain data with synthetic data to help the model "forget" specific information. An example of this approach involves generating synthetic labels for the data within $D_u$ and then combining them with the data in $D_u$ for training to accomplish unlearning. This method disentangles the impact of certain data from the model, helping to eliminate the influence of specific information while retaining the model's overall performance.

\subsubsection{Verification}
Verification methods aim to confirm whether data intended for deletion has indeed been effectively unlearned. Currently, these methods can be classified as outlined below:

\textbf{Model performance.} The most straightforward approach is to evaluate the model performance on the target client's data and test data to assess how effectively the data has been unlearned and how robustly the unlearned model is maintained. The evaluation metrics encompass accuracy, loss, and statistical errors.

\textbf{Model discrepancy.} Another approach to assess unlearning performance is by evaluating the discrepancy between the trained model and the unlearned model. This discrepancy can be measured using metrics such as Euclidean distance, KL-divergence, L2 distance, Wasserstein distance, and angle-based distance.

\textbf{Execution efficiency.} In addition, the time taken for the unlearning process, measured in terms of rounds, runtime, or speed-up ratio compared to a baseline, as well as memory consumption, can be used to evaluate the efficiency of the unlearning algorithm.

\textbf{Attack performance.} As introduced in Section \ref{sec:attacks_to_machine_unlearning}, membership inference attacks can be used to determine whether a particular data was used during the training of a model. Therefore, by executing MIA on the unlearned model over unlearned data, the attack success rate (ASR) can be used to evaluate how effectively the data has been unlearned. Poorer performance by the MIA indicates that the influence of the unlearned data on the global model has diminished. Similarly, in the context of backdoor attacks, by injecting backdoors into the unlearned data and following the unlearning procedure, effective unlearning should disrupt the relationship between the trigger pattern and the backdoor class. The ASR can also be used to evaluate how effectively the backdoor is removed by unlearning. An empirical study on these metrics can be referred to in \cite{nguyen2024empirical}.

\subsection{Federated Learning}
\label{sec:federated_learning}

\subsubsection{Overview of federated learning}

The participants involved in federated learning \cite{kairouz2021advances,mcmahan2017communication,yang2019federated} can be categorized into two categories: (i) a set of $n$ clients denoted as $\mathcal{U}={u_1,u_2,\dots,u_n}$, where each client $u_i \in \mathcal{U}$ possesses its local dataset $\mathcal{D}_i$, and (ii) a central server represented as $S$. A typical FL scheme works by repeating the following steps until training is stopped \cite{kairouz2021advances}. (i) Local model training: each FL client $u_i$ trains its model $\mathcal{M}_i$ using the local dataset $\mathcal{D}_i$. (ii) Model uploading: each FL client $u_i$ uploads its locally trained model $\mathcal{M}_i$ to the central server $S$. (iii) Model aggregation: the central server $S$ collects and aggregates clients' models to update the global model $\mathcal{M}$. (iv) Model updating: the central server $S$ updates the global model $\mathcal{M}$ and distributes it to all FL clients.

\subsubsection{Security and privacy threats in federated learning}
The revelation of a participant's local model poses a direct threat to the fundamental privacy guarantee of standard federated learning \cite{zhu2019deep}. Thus, privacy-preserving aggregation protocols \cite{bell2020secure,guo2021privacy,bonawitz2017practical,liu2023long} are essential to maintain the security and privacy of the model aggregation process in Step iii of FL. Additionally, FL is susceptible to poisoning attacks \cite{lyu2020threats} (see Section \ref{sec:attacks_to_machine_unlearning} for more details). In these attacks, malicious clients manipulate the global model by sending poisoned model updates to the server during Step ii, to affect global model performance or inject backdoors. Therefore, malicious-client detection mechanisms \cite{li2020learning, CaoF0G21, zhang2022fldetector, shen2016auror} are imperative to differentiate between malicious and benign clients.

\subsection{Federated Unlearning}
\label{sec:fu_definition}

In an FU system, the set of FL clients is represented as $U$, where each client $u_i \in U$ possesses a local dataset $D_i$. This set is categorized into two distinct subsets: $U_u$, which includes clients designated for unlearning (either entirely or partially), and $U_r$, comprising the remaining clients, with the relationship $U = U_r \cup U_u$. More specifically, for any client $u_j \in U_u$, $\bar{D}_j$ represents the data of $u_j$ to be unlearned, hence $\bar{D}_j=D_j$ signifies unlearning of the entire client $u_j$, and $\bar{D}_j \subset D_j$ indicates unlearning partial data of the client $u_j$. Now, we give the definition of federated unlearning.


\begin{definition}\textbf{(Federated Unlearning)} 
    A federated unlearning process $FU(M,U,U_u, U_r) \rightarrow \bar{M}$ is defined as a function from a global model $M$ obtained through FL $FL(U)$ trained by a set of FL clients $U$ to an unlearned model $\bar{M}$. This function considers two subsets of $U$ including the set of unlearned client $U_u \subset U$ where each $u_j \in U_u$ posses its unlearned dataset $\bar{D}_j$, and the set of remaining client $U_r \subset U$. The goal is to ensure that the unlearned global model $\bar{M}$ maintains performance comparable to a retrained model $\hat{M}$ trained by $U_r \cup U_u$ where each $u_i \in U_r$ posses $D_i$ and each $u_j \in U_u$ posses $D_j \backslash \bar{D}_j$.
\end{definition}
Building on the definition of federated unlearning outlined above, we specify the targets of the unlearning process within an FL setting as follows.
\begin{definition}\textbf{(Model consistency)}
    For a given set of samples $X$, let $\bar{Y}$ be the predicted results produced from the unlearned global FL model $\bar{M}$, and $\hat{Y}$ be the predicted results from a retrained global FL model $\hat{M}$. Then, the unlearning process $FU(M,U,U_u, U_r)$ is considered to provide full consistency if $\bar{Y} = \hat{Y}$. The target regarding model consistency is to make the performance of the unlearned model $\bar{M}$ as much as similar to that of $\hat{M}$.
\end{definition}
\begin{definition}\textbf{(Unlearning efficiency)}
    For a retrained model $\hat{M}$ and an unlearned model $\bar{M}$ obtained from $FU(M,U,U_u, U_r)$ with full consistency, the target regarding unlearning efficiency is to make the cost of $FU(M,U,U_u, U_r)$ as much less than the cost of obtaining the retrained model $\hat{M}$.
\end{definition}
\begin{definition}\textbf{(Privacy preservation)}
    For a federated learning process $FL(U)$ followed by a federated unlearning process $FU(M,U,U_u, U_r)$, the target regarding privacy preservation is to ensure that the additional information leakage caused by $FU(M,U,U_u, U_r)$, beyond what is leaked through $FL(U)$, is kept as minimal as possible.
\end{definition}
\begin{definition}\textbf{(Certified removal)}
    For any participant, whether an FL client or server, initiating the unlearning request, the target regarding certified removal is to establish a function $V(\cdot)$, which serves to confirm that the unlearning process $FU(M,U,U_u, U_r)$ has been carried out in accordance with the request made by that participant.
\end{definition}

\subsection{Attacks to ML Models}
\label{sec:attacks_to_machine_unlearning}

As mentioned earlier, attacks on ML models can serve as a means to verify the effectiveness of unlearning. Specifically, these attacks can help determine whether the data related to the target client has been successfully unlearned. In this section, we will primarily introduce the two most widely adopted attack methods that are utilized for unlearning verification:

\subsubsection{Membership inference attacks (MIA)}
First proposed by Shokri \textit{et al.} \cite{shokri2017membership}, the fundamental idea behind MIA is to determine whether a particular record was used during the training of a target model. This is predicated on the observation that data samples present in the training set will lead the model to produce outputs with higher confidence scores. Consequently, an adversary can train a separate model for binary classification, designating outputs as either ``member" (indicating that the data was part of the training set) or ``non-member" (indicating that the data was not part of the training set). This potential to distinguish between member and non-member records poses a threat to data privacy. 
Remarkably, MIA does not require knowledge of the target model's specific architecture or the distribution of its training data. Relying on the shadow models, a series of shadow training datasets $D'_1,\cdots,D'_k$ and disjointed shadow test datasets $T'_1,\cdots,T'_k$ can be synthesized to mimic the behavior of the target model so as to train the attack model. Evaluating the unlearning effectiveness can be achieved by performing MIA on the unlearned model with the unlearned data. The ASR serves as an indicator of how well the data has been unlearned. A decrease in the MIA's performance suggests that the influence of the unlearned data on the global model has been successfully reduced.

\subsubsection{Backdoor attacks (BA)}
Backdoor attacks embed a distinct pattern or ``trigger" into portions of the training data \cite{li2022backdoor}. The trigger can be a small patch or sticker that is visible to humans \cite{gu2017badnets,liu2018trojaning}, or the value perturbation of benign samples indistinguishable from human inspection \cite{li2020invisible,bagdasaryan2021blind,saha2020hidden}. When the model is subsequently trained or fine-tuned on this, its behavior remains typical for standard inputs. Yet, upon detecting an input that contains this covert trigger, the model will yield malicious behaviors that align with the attacker's intentions. Backdoor attacks are particularly concerning because they can remain dormant and undetected until the attacker chooses to exploit them. In the context of unlearning, injecting backdoors into the unlearned data and then applying the unlearning procedure should effectively disrupt the relationship between the trigger pattern and the backdoor class. The ASR can be utilized to assess the effectiveness of the unlearning process in removing the backdoor. A lower ASR would indicate that the backdoor has been successfully eliminated.

\section{Federated Unlearning Methods}
\label{sec:federated_unlearning_methods}

\begin{figure*}[htbp!]
    \centering
    \includegraphics[width=0.8\linewidth]{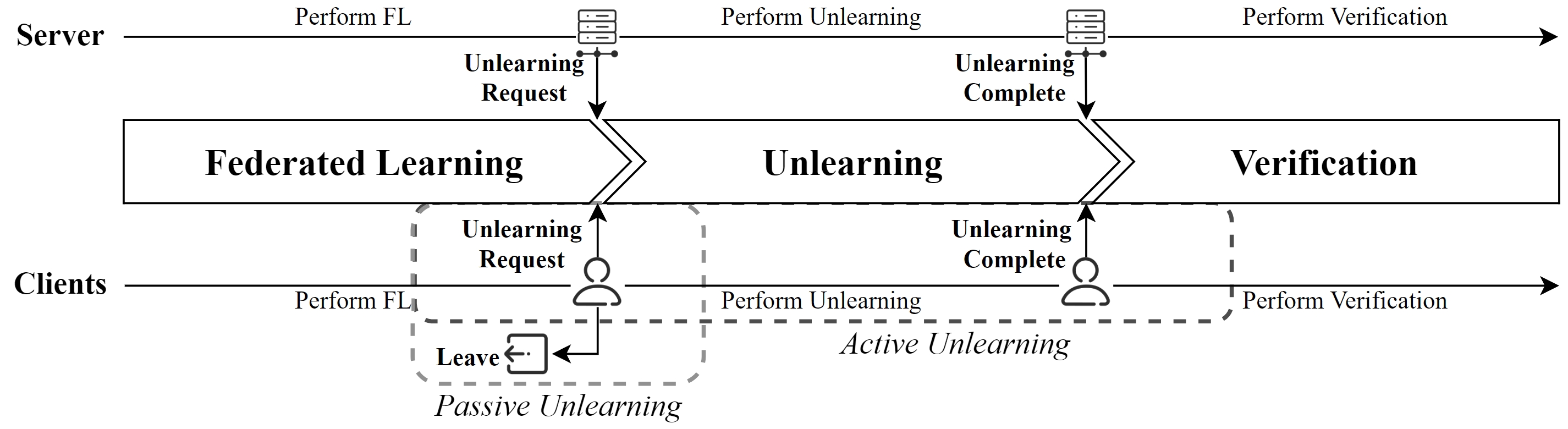}
    \caption{An unified federated unlearning workflow. This workflow outlines the timeline for learning, unlearning, and verification. When the FU system receives an unlearning request, it can follow either the passive unlearning approach, where the target client exits the system immediately, or the ``Active unlearning" approach, where the target client chooses to stay and participate in the unlearning process. Unlearning requests can be initiated by either the unlearned client or the server for various purposes. Furthermore, the unlearning and verification roles can be performed by the server, the target clients, the remaining clients, or a combination of both.}
    \label{fig:fu-workflow}
\end{figure*}

\begin{figure*}[htbp!]
    \centering
    \includegraphics[width=1\linewidth]{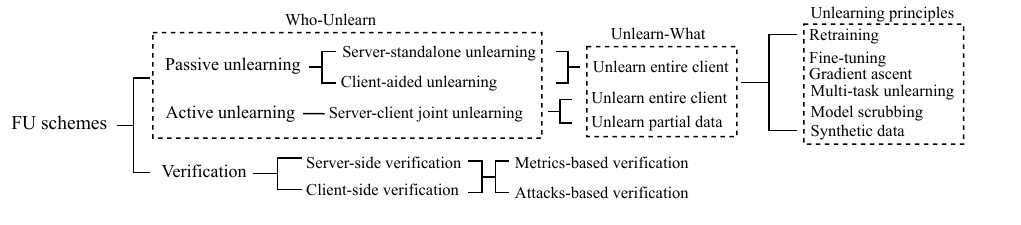}
    \caption{Taxonomy of federated unlearning schemes.}
    \label{fig:fu-taxonomy}
\end{figure*}

In this section, we introduce a unified federated unlearning workflow, as illustrated in Figure \ref{fig:fu-workflow}, serving as the basis for a novel taxonomy of existing FU techniques. This workflow defines the timeline for learning, unlearning, and verification. When the FU system receives an unlearning request, it can either allow the target client to exit the system immediately, referred to as ``Passive unlearning," or the target client can choose to stay and participate in the unlearning process, referred to as ``Active unlearning."
Note that some unlearned clients may simultaneously initiate the unlearning request and transmit information to the server, while others may not engage in the unlearning process but remain solely for verification. We categorize these FU schemes as passive unlearning as well. The taxonomy can be found in Figure \ref{fig:fu-taxonomy} and the summary of the existing FU works can be found in Table \ref{tab:fu_summary_table}.

\subsection{Passive unlearning}

Passive unlearning signifies that the target client does not stay within the FU system to participate in the unlearning process, which typically involves a series of computational iterations. Instead, the remaining participants, including the central server, the remaining FL clients, or both, carry out the unlearning algorithms. In this case, passive unlearning unlearns the entire client instead of partial data. In the scenario of (i) server-standalone unlearning, historical information such as gradients and global models is stored, enabling the server to eliminate the influence of the unlearned client using various methods. In the scenario of (ii) client-aided unlearning, the standard FL workflow is followed, with iterative refinements of the global model achieved by aggregating improved information from the remaining clients. Note that for passive unlearning, methods like local retraining, local fine-tuning, and multi-task unlearning are no longer applicable. This is because these FU algorithms rely on direct access to both the unlearned data and the remaining data of the target client, which becomes inaccessible when the entire client must be removed.

\subsubsection{Server-standalone unlearning}

As previously mentioned, standalone server unlearning typically depends on the utilization of stored historical data, which may include gradients, global models \cite{wu2023unlearning,wu2022distill,zhang2023fedrecovery,guo2023fast,jiang2024towards,huynh2024fast}, contribution information \cite{zhang2023fedrecovery}, or intermediate information necessary for constructing a random forest \cite{liu2021revfrf}. This category necessitates a significant amount of memory on the server, potentially limiting its practical application in large-scale FL systems with complex ML models.

In the case of FedRecovery \cite{zhang2023fedrecovery}, the server retains historical data from all clients and quantifies their contributions in each round based on gradient residuals. When a target client requests to leave, the server systematically removes its contributions from all FL rounds through a fine-tuning process. Based on FedRecovery \cite{zhang2023fedrecovery}, Crab \cite{jiang2024towards} achieves a more efficient recovery based on (i) selective historical information rather than all historical information and (ii) a historical model that has not been significantly affected by malicious clients rather than the initial model. Additional constraints can be introduced to further guide the recovery process, such as a penalty term based on projected gradients \cite{shao2024federated,fu2024client}, randomly initialized degradation models \cite{zhao2023federated}, and estimated skew \cite{huynh2024fast}. The approach of eliminating the contribution of the target client is more straightforward in \cite{wu2023unlearning, wu2022distill}, where the server directly averages the models of the remaining clients. Strategic retraining based on the change of sampling probability is adopted for fast and efficient recovery \cite{tao2024communication}. To mitigate the potential decrease in accuracy due to the averaging process in the averaged model, knowledge distillation is employed. This technique facilitates the transfer of information from the trained model to the unlearned model, helping to preserve performance. Consequently, these designs adhere to a multi-task unlearning approach. In VERIFI \cite{gao2022verifi}, after receiving the gradients from all clients, including those from the target client, are uploaded to the server. The server then amplifies the gradients from the remaining clients and diminishes the gradients of the target client, to reduce the impact of target clients, hence achieving unlearning.

\begin{figure}[htbp!]
    \centering
    \includegraphics[width=0.8\linewidth]{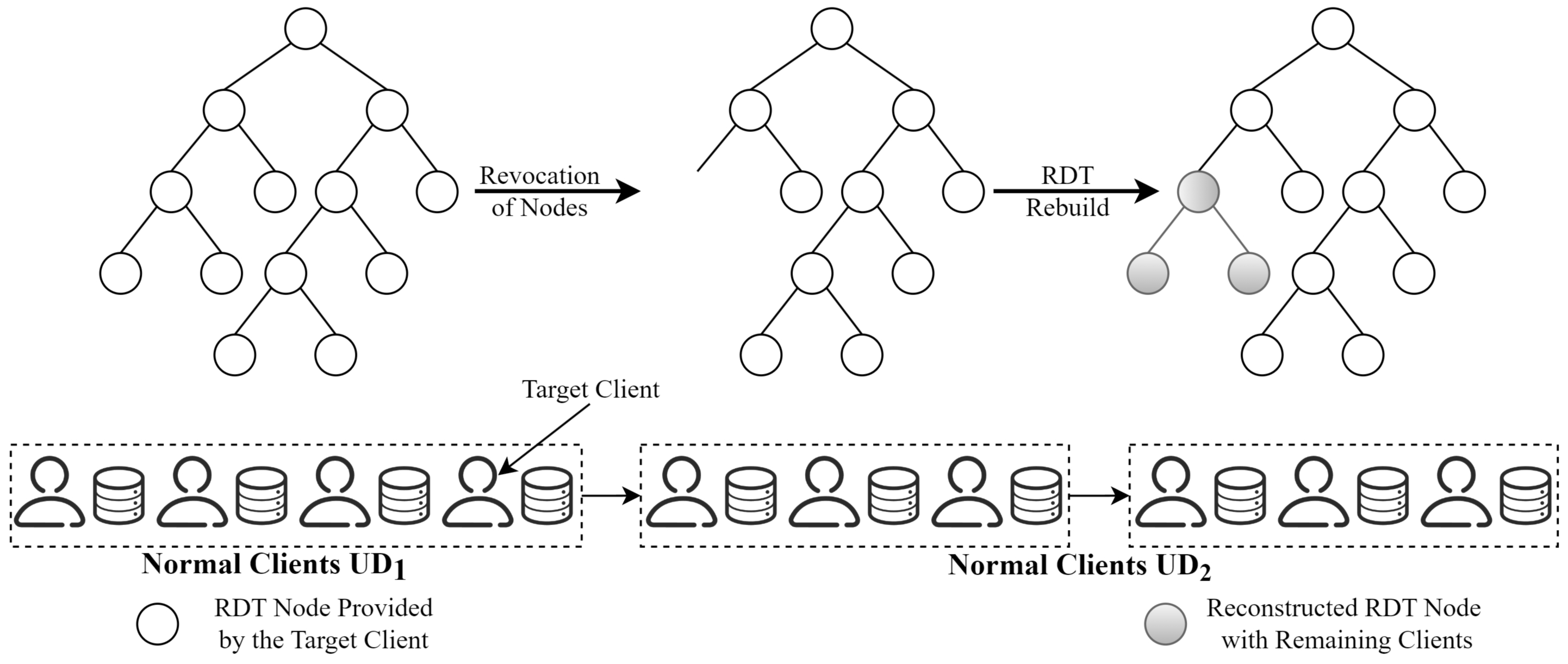}
    \caption{An illustration of RevFRF \cite{liu2021revfrf}. To remove a target client, the server first identifies nodes affected by the target client and subsequently eliminates these affected nodes until reaching the leaf node. Following this, the server reconstructs the affected branches through a retraining process, based on previously stored intermediate information.}
    \label{fig:RevFRF}
\end{figure}

Apart from the above-mentioned works, RevFRF \cite{liu2021revfrf}, as shown in Figure \ref{fig:RevFRF}, focuses on federated random forest training. To remove a client, the server first identifies nodes affected by the target client and subsequently eliminates these affected nodes until reaching the leaf node. Subsequently, the server reconstructs the affected branches using previously stored intermediate information, rather than instructing the remaining clients to undergo retraining. In particular, in the worst-case scenario where the revoked node is the root node, the server has to reconstruct the entire random decision tree. Unlearning within a federated clustering setting is explored in SCMA \cite{pan2022machine}, where each client maintains a vector to denote its local clustering result. These vectors are then aggregated by the server to form a global clustering outcome. Eliminating a client is straightforward by assigning a zero vector to the unlearned client and then re-aggregating all vectors.

\textbf{Limitations.} Server-standalone unlearning, which relies on historical data stored on the server, lacks real-time input from remaining clients during the unlearning process. This limitation may result in slightly lower unlearning performance compared to client-aided unlearning. This characteristic could impede the applicability of server-standalone unlearning in complex ML models or Non-IID FL settings, where there is a notable bias, affecting the overall efficacy of the unlearning process. Furthermore, server-standalone unlearning may lack responsiveness to changes in data and client behavior, as it solely relies on historical information. This can limit its adaptability in dynamic environments where real-time data and client interactions are crucial.

\subsubsection{Client-aided unlearning}
\label{sec:joint_learning}
Unlearning performed by the server and remaining clients typically offers greater potential compared to standalone server unlearning. This is because the remaining clients contribute valuable information about the remaining data, which enables the server to enhance its unlearning process. In this context, the server may or may not have access to historical information.

\begin{figure}[htbp!]
    \centering
    \includegraphics[width=0.7\linewidth]{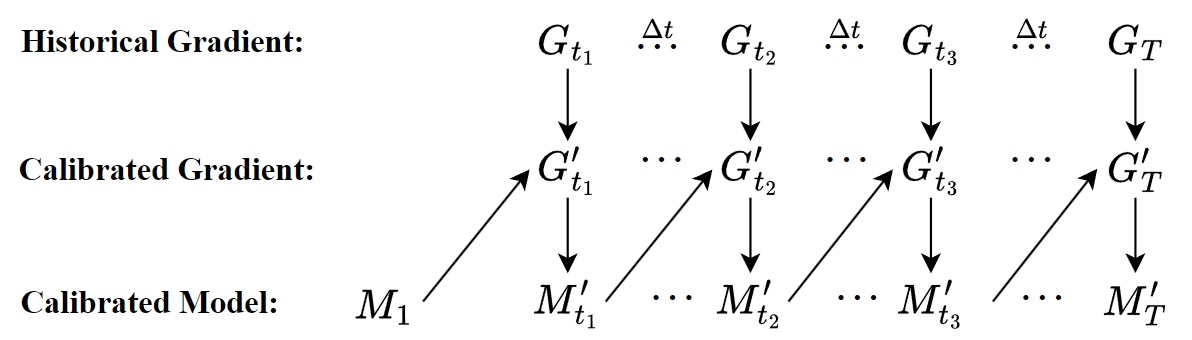}
    \caption{An illustration of FedEraser \cite{liu2021federaser}. The server stores clients' gradients at intervals of every $\Delta t$ rounds. Using an iterative approach, for a given round $t_i$, the server computes calibrated gradients $G_{t_i}^{\prime}$ based on historical gradients $G_{t_i}$ and the calibrated model $M_{t_{i-1}}^{\prime}$.}
    \label{fig:FedEraser}
\end{figure}

This research direction is arguably pioneered by the design of FedEraser \cite{liu2021federaser}, as shown in Figure \ref{fig:FedEraser}. The core concept of FedEraser is that the current global model can be reconstructed using only the initial model and historical clients' gradients at each round. Consequently, unlearning boils down to eliminating the influence of the target client on the historical gradients, i.e., calibrating historical gradients. To achieve this goal, for a historical FL round $i$ with the stored gradients $G_i$ and a calibrated global model $M_{i-1}$ for the previous round, each remaining client $u_j$ calculates a local calibration direction $c_i^j$ based on its local data $D_j$ and $M_{i-1}$. The local calibration directions are then aggregated by the server to derive a global calibration direction $c_i$, which enables the server to calculate calibrated historical gradients $G_i^{\prime}$ and to obtain a calibrated global model $M_i$ via $G_i^{\prime}$. This iterative process continues round-by-round until all historical gradients are successfully calibrated, resulting in the server obtaining a final calibrated global model, eliminating the influence of the target client. To enhance unlearning efficiency, the server stores clients' gradients at intervals of every $\Delta t$ rounds, leading to a trade-off between unlearning performance and resource consumption in terms of memory and computation. Improve upon FedEraser, Crab \cite{jiang2024towards} and Fast-FedUL \cite{huynh2024fast} optimize storage efficiency by selectively storing important gradients, while Sharding Eraser \cite{lin2024scalable} compress storage using coding-based techniques. A similar idea is adopted in \cite{wang2024forget} focusing on ranking tasks instead of classification tasks. Building upon the unlearning concept introduced in FedEraser, an efficiency-enhancing technique is employed in FRU \cite{yuan2023federated} for federated recommendations. In FRU, only the important updates to clients' item embeddings are stored. In line with FedEraser \cite{liu2021federaser} and FRU \cite{yuan2023federated}, FedRecover \cite{cao2023fedrecover} also entails the storage of historical gradients and global models. In FedRecover, to prevent the remaining clients from computing exact model updates for fine-tuning, which can lead to significant computational overhead, the server calculates updates for the remaining clients using historical gradients and global models, as described below:
$$
\boldsymbol{g}_t^i=\overline{\boldsymbol{g}}_t^i+\mathbf{H}_t^i\left(\hat{\boldsymbol{w}}_t-\overline{\boldsymbol{w}}_t\right)
$$

where $\mathbf{H}_t^i=\int_0^1 \mathbf{H}\left(\overline{\boldsymbol{w}}_t+z\left(\hat{\boldsymbol{w}}_t-\overline{\boldsymbol{w}}_t\right)\right) d z$ is an integrated Hessian matrix for the $i$th client in the $t$th round. Denote the global-model difference in the $t$th round as $\Delta \boldsymbol{w}_t=\hat{\boldsymbol{w}}_t-\overline{\boldsymbol{w}}_t$ and the model-update difference of the $i$th client in the $t$th round as $\Delta \boldsymbol{g}_t^i=\boldsymbol{g}_t^i-\overline{\boldsymbol{g}}_t^i$. The Hessian matrix $\tilde{\boldsymbol{H}}_t^i$ for the $i$th client in the $t$th round can be approximated as
$$
\tilde{\boldsymbol{H}}_t^i=\text{L-BFGS}\left(\Delta \boldsymbol{W}_t, \Delta \boldsymbol{G}_t^i\right)
$$

where $\Delta \boldsymbol{W}_t=\left[\Delta \boldsymbol{w}_{b_1}, \Delta \boldsymbol{w}_{b_2}, \cdots, \Delta \boldsymbol{w}_{b_s}\right]$ and $\Delta \boldsymbol{G}_t^i=\left[\Delta \boldsymbol{g}_{b_1}^i, \Delta \boldsymbol{g}_{b_2}^i, \cdots, \Delta \boldsymbol{g}_{b_s}^i\right]$ are L-BFGS buffers \cite{nocedal1980updating} maintained by the server. Nonetheless, these approximations introduce estimation errors over rounds. Therefore, the remaining clients are periodically tasked with computing their exact model updates to correct these approximations, based on an adaptive abnormality threshold. A more straightforward retraining-based method is employed in SIFU \cite{fraboni2022sequential}. In SIFU, the fundamental concept of unlearning is to identify the most recent global model using a bounded sensitivity metric calculated from historical contributions. Subsequently, the remaining clients retrain based on the identified model.

\begin{figure}[htbp!]
    \centering
    \includegraphics[width=0.7\linewidth]{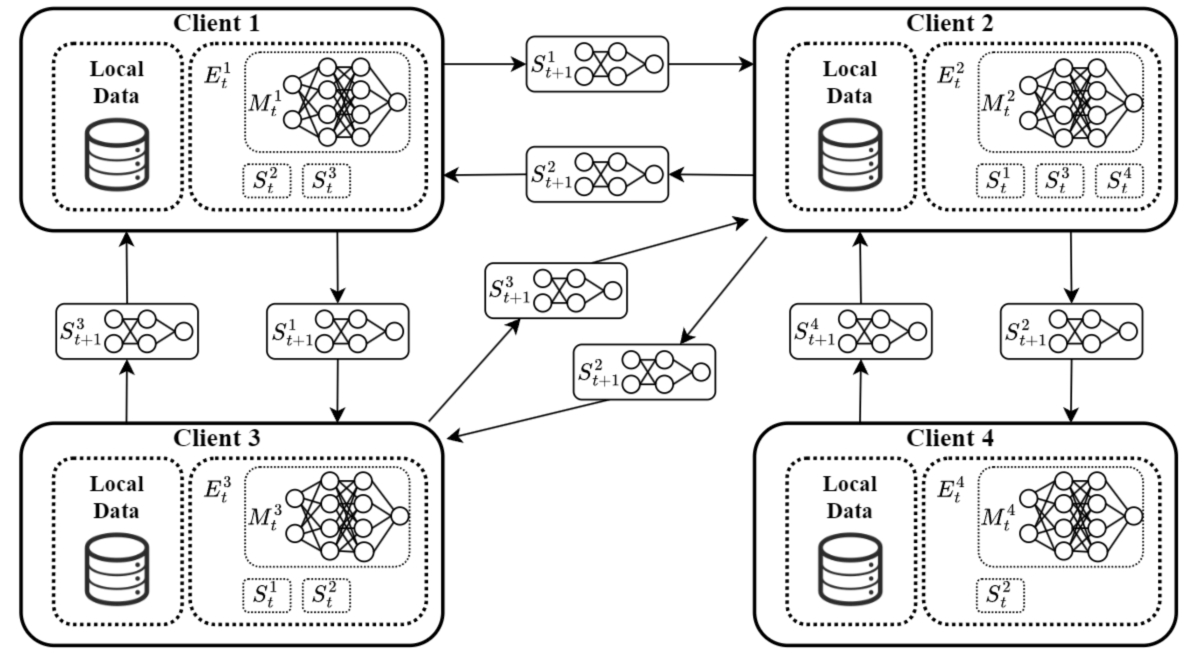}
    \caption{An illustration of HDUS \cite{ye2023heterogeneous}. Operating without a central server, each client possesses their own neighboring distilled models, referred to as seed models. When a client requests to leave, the adjacent clients simply delete the seed model of the unlearned client. For predictions, an ensemble learning method is employed to combine the outputs of the primary model with those of the seed models.}
    \label{fig:HDUS}
\end{figure}

Other approaches concentrate on unlearning without reliance on historical updates. In SFU \cite{li2023subspace}, upon receiving an unlearning request, the server refines the global model using gradient information provided by the target client and representation matrix information provided by other clients. In KNOT \cite{su2023asynchronous}, clients are grouped into clusters based on their training time and model sparsity. Clients within the same cluster collectively aggregate and update the global model asynchronously. When a client requests to leave, only clients within the same cluster require retraining. A similar structure is adopted in FedCIO \cite{qiu2023fedcio} where clients are clustered according to their data distribution. Differing from conventional FL systems, HDUS \cite{ye2023heterogeneous} operates without a central server, as shown in Figure \ref{fig:HDUS}. Instead, each client possesses their own neighboring distilled models, referred to as seed models. When a client requests to leave, the adjacent clients simply delete the seed model of the unlearned client. For predictions, an ensemble learning method is employed to combine the outputs of the primary model with those of the seed models. Incentive mechanisms along with game theoretical analysis in FU systems is presented in \cite{ding2023incentive,lin2024incentive,ding2023strategic}.

\begin{table}[htbp!]
\caption{A Summary of Passive FU schemes}
\label{tab:fu_summary_table}
\begin{centering}
\begin{tabular}{|>{\centering\arraybackslash}m{0.3cm}|>{\centering}m{0.5cm}|>{\centering\arraybackslash}m{1.3cm}|>{\centering}m{1.3cm}|>{\centering}m{1.3cm}|>{\centering}m{4cm}|>{\centering}m{1cm}|>{\centering}m{2.4cm}|}
\hline
 \cellcolor{mygray}  \textbf{\noun{}} &  \cellcolor{mygray} \textbf{\noun{Ref.}}&  \cellcolor{mygray} \textbf{\noun{Who-Unlearn}}&  \cellcolor{mygray} \textbf{\noun{Unlearn-What}}&  \cellcolor{mygray} \textbf{\noun{Principle}}  & 
 \cellcolor{mygray} \textbf{\noun{Method}} &
 \cellcolor{mygray} \textbf{\noun{Verifier}} &  \cellcolor{mygray}\textbf{\noun{Verify Method}} \tabularnewline
\hline
\hline

\parbox[t]{2mm}{\multirow{6}{*}{\rotatebox[origin=c]{90}{ \hspace{-14 cm} Passive Unlearning}}}

&\cite{zhang2023fedrecovery}\\ \cite{jiang2024towards}& Server & Target Client & Fine-tuning & Iteratively remove the contributions of the target client evaluated based on its historical gradient residuals.& NA & Accuracy-based metrics, unlearning time, MIA
\tabularnewline \cline{2-8}

&\cite{cao2023fedrecover}& Server \& Remaining clients & Target client & Model scrubbing & The server scrubs the model iteratively based on the estimation over historical gradients and global models, while the remaining clients periodically participate to eliminate accumulated estimation errors. & NA & Test error rate, backdoor attack, average computation/communication costs saving
\tabularnewline \cline{2-8}

&\cite{liu2021revfrf}& Server & Target client & Retraining & Remove nodes affected by the target client until reaching the leaf node. Then reconstruct the affected branches based on previously stored intermediate information. & NA & Accuracy-based metrics
\tabularnewline \cline{2-8}

&\makecell{\cite{wu2023unlearning} \\ \cite{wu2022distill}}& Server & Target client & Multi-task unlearning & Unlearn by directly averaging the models of remaining clients, while avoiding forgetting by optimizing the knowledge distillation loss between unlearned model and previous global model.& NA & Accuracy-based metrics, backdoor attack
\tabularnewline \cline{2-8}

&\cite{su2023asynchronous}& Sever \& Target client \& Some remaining clients & Target client & Retraining & Divide clients into clusters for asynchronous aggregation. To unlearn some data, only clients in the same cluster are retrained. & Sever & Validation accuracy, deviation across recent validation accuracies
\tabularnewline \cline{2-8}

&\cite{li2023subspace}& Server \& All clients & Target client & Gradient ascent & The server refines the global model with gradient ascent in a subspace based on the gradient provided by the target client and the representation matrix provided by the remaining clients. & NA & Backdoor attack
\tabularnewline \cline{2-8}

&\cite{fraboni2022sequential}& Server \& Remaining clients & Target client & Retraining & Find a historical global model based on a bounded sensitivity metric calculated based on clients' historical contributions, from where the remaining clients retrain. & NA & Number of retraining rounds, accuracy-based metrics
\tabularnewline \cline{2-8}

\hline
\hline

\end{tabular}
\par\end{centering}
\end{table}

\begin{table}[htbp!]
\caption{A Summary of Passive FU schemes (continued).}
\label{tab:fu_summary_table_continue_1}
\begin{centering}
\begin{tabular}{|>{\centering\arraybackslash}m{0.3cm}|>{\centering}m{0.5cm}|>{\centering\arraybackslash}m{1.3cm}|>{\centering}m{1.3cm}|>{\centering}m{1.3cm}|>{\centering}m{4cm}|>{\centering}m{1cm}|>{\centering}m{2.4cm}|}
\hline
 \cellcolor{mygray}  \textbf{\noun{}} &  \cellcolor{mygray} \textbf{\noun{Ref.}}&  \cellcolor{mygray} \textbf{\noun{Who-Unlearn}}&  \cellcolor{mygray} \textbf{\noun{Unlearn-What}}&  \cellcolor{mygray} \textbf{\noun{Principle}}  & 
 \cellcolor{mygray} \textbf{\noun{Method}} &
 \cellcolor{mygray} \textbf{\noun{Verifier}} &  \cellcolor{mygray}\textbf{\noun{Verify Method}} \tabularnewline
\hline
\hline

\parbox[t]{2 mm}{\multirow{3}{*}{\rotatebox[origin=c]{90}{ \hspace{-14 cm} Passive Unlearning}}}

&\cite{yuan2023federated}&Server \& Remaining clients &Target client & Fine-tuning &Iteratively and selectively calibrate historical gradients to reconstruct the calibrated global model.  &NA &Backdoor attack
\tabularnewline \cline{2-8}

&\cite{ye2023heterogeneous}& Remaining clients& Target Client& Model scrubbing & Each client retains neighboring distilled models, and predictions are obtained through an ensemble of the main model and seed model. To unlearn a target client, simply delete the seed model associated with that target client. & NA & Accuracy-based metrics
\tabularnewline \cline{2-8}

&\cite{liu2021federaser}& Server \& Remaining clients & Target client & Fine-tuning &  Iteratively calibrate historical gradients to reconstruct the calibrated global model. & NA & Metrics, parameter deviation, MIA 
\tabularnewline \cline{2-8}

&\cite{pan2022machine}& Server & Target client & Fine-tuning & Server aggregate the vectors from remaining clients representing their local clustering result. & Server & Global model convergence
\tabularnewline \cline{2-8}

&\cite{gao2022verifi}& Server& Target client& Fine-tuning & The server then amplifies the gradients from the remaining clients and reduces the gradients of the target client. & Target client & Accuracy-based metrics
\tabularnewline \cline{2-8}

&\cite{lin2024scalable}& Server \& Remaining clients & Target client & Retraining & Retraining based on isolated shard and coded computing& NA & Accuracy, time, storage, MIA
\tabularnewline \cline{2-8}

&\cite{wang2024forget}& Remaining clients & Target client & Fine-tuning & Iteratively calibrate historical gradients to reconstruct the calibrated global model. & NA & Backdoor attack
\tabularnewline \cline{2-8}

&\cite{shao2024federated} \\ \cite{fu2024client}& Server & Target client & Fine-tuning & Calibrate historical gradients with penalty term based on projected gradients. & NA & Accuracy-based metrics, backdoor attack
\tabularnewline \cline{2-8}

&\cite{guo2023fast}& Server & Target client & Fine-tuning &  Fine-tuning the model by subtracting target model updates. & NA & Accuracy-based metrics, time, CPU usage, memory
\tabularnewline \cline{2-8}

&\cite{huynh2024fast}& Server & Target client & Fine-tuning &  Calibrate historical gradients with guidance of estimated skew. & NA & Accuracy, backdoor attack
\tabularnewline \cline{2-8}

&\cite{tao2024communication}& Server & Target client \& Partial data & Retraining & Strategic retraining based on the change of sampling probability. & NA & Accuracy, time, MIA
\tabularnewline \cline{2-8}

\hline
\hline

\end{tabular}
\par\end{centering}
\end{table}

Client-aided unlearning inherently depends on the involvement of remaining clients and their updates, which can be a vulnerability in dynamic environments where client participation fluctuates. Additionally, this unlearning method can be slow, as it relies on all remaining clients, often resource-constrained devices, and is limited by the bandwidth of the FL system. This could lead to inefficiencies, particularly in cross-device scenarios with frequent client turnover or limited system resources.

\subsection{Active unlearning}
``Active unlearning" denotes that the target client actively engages in the unlearning process and then has the option to either stay or leave, with or without verification (see Section \ref{sec:verification} for more details on verification mechanisms). Given the direct access the target client possesses to the data to be unlearned, this approach exhibits greater potential as indicated by existing research. 

\subsubsection{Unlearn partial data}
To unlearn the partial data of the target client, retraining is the most straightforward approach. To mitigate the computational cost of starting from scratch with retraining, one solution is to roll back the global model to a state where it has not been significantly influenced by the target client. From this point, all FL clients can conduct the retraining process. For instance, in Exact-Fun \cite{xiongexact}, where FL models are quantized, when a client requests to leave, the client calculates a new model based on the remaining data. If the original model matches the new quantized model, signifying that the removal has no impact, the FL model remains unchanged. Otherwise, retraining is required to eliminate the influence of the unlearned data. In ViFLa \cite{fan2022fast}, which is essentially a machine unlearning scheme, training samples are segmented into different groups, with each group representing an FL client. Hence, ViFLa can simulate an FU process in this context. The local model is trained using ring-based SQ-learning for LSTM, and weighted aggregation is determined by KL-attention scores. The historical model parameters represented by states over a ring are stored. To unlearn partial data, each client removes unlearned data and computes the new updates. Based on these new updates, the server identifies a previous state from which the remaining clients continue their training. A similar concept of identifying the optimal previous state for retraining is also present in SIFU \cite{fraboni2022sequential}, as discussed earlier in Section \ref{sec:joint_learning}. In SCMA \cite{pan2022machine}, a straightforward approach to unlearning partial data involves naive retraining. Each client maintains a vector representing its local clustering result, and these vectors are aggregated by the server to create a global clustering outcome. To unlearn partial data, SCMA entails each client calculating a new local vector, i.e., retraining, and then re-aggregating all vectors.

Fine-tuning and multi-task learning are popular approaches for FU as well. As an example, in FRAMU \cite{shaik2023framu}, the server aggregates fine-tuned local models and attention scores. Using these scores, it filters out irrelevant data points and updates the global model. The attention scores are acquired through local reinforcement learning applied to dynamic data. In FedLU \cite{zhu2023heterogeneous}, designed for FL over knowledge graphs where embeddings are aggregated instead of gradients as in standard FL, the unlearning of partial data is accomplished through iterative optimization of local embeddings. This process follows the multi-task unlearning concept, involving unlearning over the local model and learning over the global model. In FedME$^2$ \cite{xia2023fedme}, clients engage in multi-task learning to optimize the loss of the local model, the loss from an MIA-like evaluation model, and a penalty term that accounts for the difference between the local model and the global model. In \cite{wang2024goldfish}, unlearning is conducted by optimizing model performance on the remaining dataset while considering bias caused by the unlearned data.

Model scrubbing-based methods are commonly employed for unlearning partial data. In \cite{liu2022right}, the model scrubbing technique is applied to the target client to locally unlearn the partial data, involving Hessian matrix computations, with enhanced computational efficiency through an approximate diagonal empirical Fisher Information Matrix (FIM). In Forsaken \cite{liu2020learn}, dummy gradients are computed to align the confidence vector of the unlearned model with that of a perfectly unlearned model. Forsaken+ \cite{ma2022learn} minimizes the distance between the posteriors of the data to be forgotten and those of non-member data for unlearning. FedAU \cite{gu2024unlearning} relies on the linear combination to approximate the unlearned model utilizing a pre-computed auxiliary model during the learning process. \cite{gu2024ferrari} focus on feature unlearning by minimizing local feature sensitivity through model scrubbing. A similar approach of local unlearning followed by aggregation is described in CONFUSE \cite{meerzaconfuse} for multi-task unlearning at different levels. FFMU \cite{che2023fast} treats data removals as perturbations on the dataset, employing random smoothing (RS) \cite{cohen2019certified} to obtain a smoother model to simulate an unlearning process. In particular, FFMU aligns with the fundamental idea presented in PCMU \cite{zhang2022prompt}, which involves randomized gradient smoothing combined with gradient quantization as follows.
$$
S(\bar{G})=\underset{c \in\{-1,0,1\}}{\operatorname{argmax}} \underset{\mathcal{D}}{\mathbb{P}}(Q(\bar{G}+\varepsilon)=c)
$$
where $\mathcal{D}=\mathcal{N}\left(0, \sigma^2 I\right)$ is a Gaussian distribution, $Q$ is a gradient quantization to map each dimension of the continuous gradient $G(x, y) \in \mathbb{R}^T$ over a discrete three-class space $\{-1,0,1\}$, for mimicking the classification in the randomized smoothing for certified robustness. $S$ is a smooth version of $Q$, and returns whichever gradient classes $Q^t$ is most likely to return when $\bar{G}$ is perturbed by noise $\varepsilon$. Extending FFMU from PCMU poses a challenge due to FL's privacy requirements, limiting server access to clients' local training data and affecting the certified data removal radius and budget in the global model. Therefore, by leveraging the theory of nonlinear functional analysis, the local MU models $g(x;q)$ in FFMU are reformulated as output functions of a Nemytskii operator $O(q)(x)$ where $q=Q(\bar{G})+\varepsilon$. In this way, the global unlearned model with bounded errors can maintain the certified radius and budget of data removals of the local unlearned models within a distance $\frac{(K-1)Cd}{\sqrt{2\pi}K\delta}$ (see \cite{che2023fast} for more details).

\begin{figure}[htbp!]
    \centering
    \includegraphics[width=0.4\linewidth]{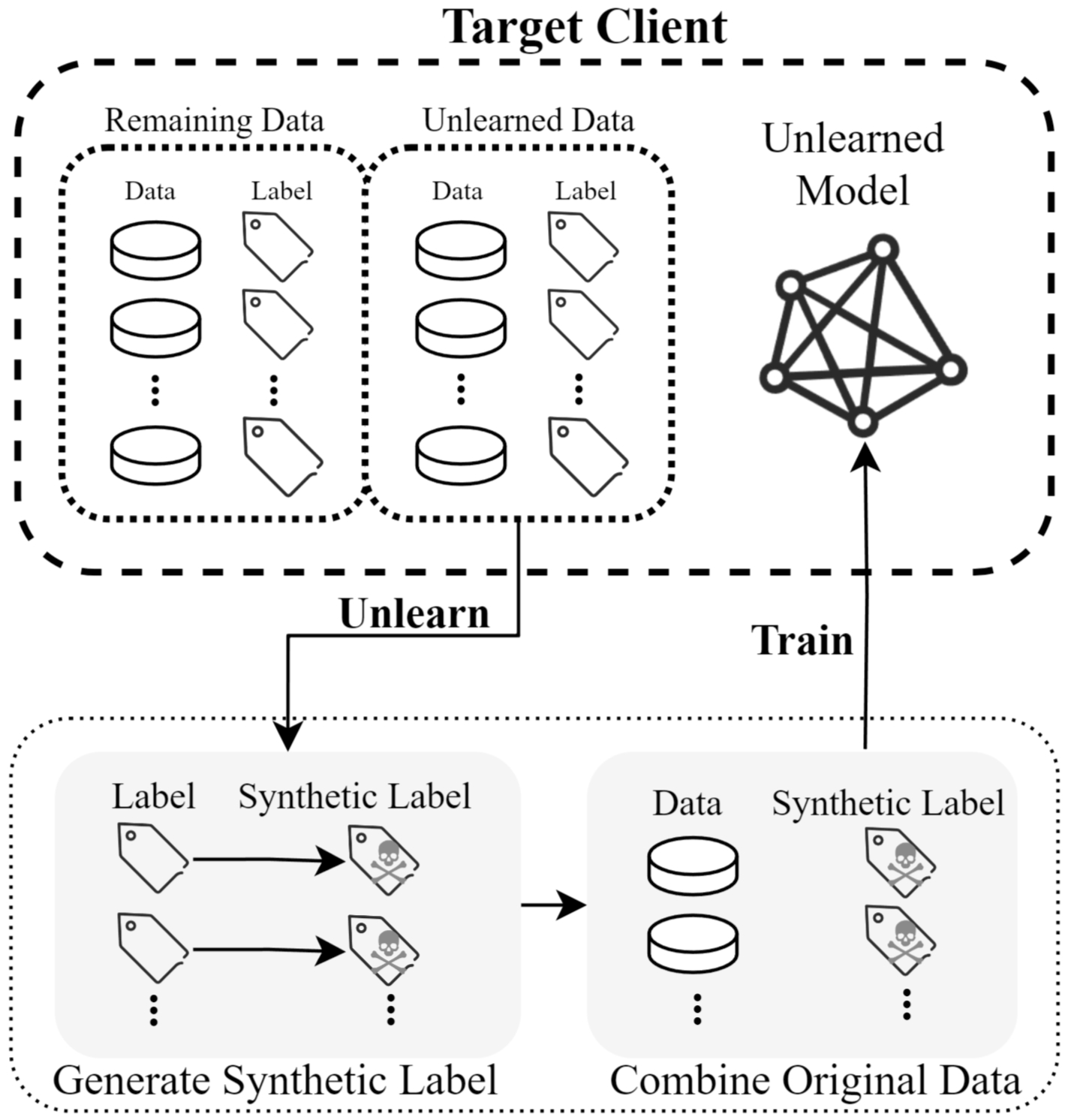}
    \caption{An illustration of FedAF \cite{li2023federated}. Synthetic labels are generated for the data to be unlearned. A trusted third party creates random teacher models, and ensemble predictions from these models to provide synthetic labels for the unlearned data. Training is then conducted using this data with synthetic labels to achieve unlearning.}
    \label{fig:FedAF}
\end{figure}

Some FU approaches are proposed with the use of synthetic data and gradient ascent principles. For instance, in UKRL\cite{xu2023revocation}, unlearning is conducted by training on perturbed unlearned data. In FedAF \cite{li2023federated}, shown in Figure \ref{fig:FedAF}, synthetic labels are generated for the data to be unlearned. A trusted third party creates random teacher models, and ensemble predictions from these models to provide synthetic labels for the unlearned data. Training is then conducted using this data with synthetic labels to achieve unlearning. Note that a multi-task approach is also employed in FedAF, where an additional task is introduced to retain the memory of the remaining data. In another work \cite{alam2023get}, which focuses on how adversaries can stealthily perform backdoor unlearning to evade server detection, the gradient ascent method is employed. The loss from the local benign dataset is utilized to constrain unbounded losses during the unlearning process based on gradient ascent.

Unlearning is also applicable in Bayesian federated learning systems. In Forget-SVGD \cite{gong2022forget}, when a client requests to leave, the target client leverages the remaining data to compute the posterior probability, approximated through variational inference (VI), and performs an extra FL round for model updating. Similar unlearning methods are employed in \cite{gong2022compressed, gong2023bayesian}. BVIFU \cite{gong2021bayesian} shares a core concept with Forget-SVGD, employing exponential family distributions in VI to approximate posterior probability. Unlike retraining-based unlearning in Bayesian FL, BFU \cite{wang2023bfu} introduces a multi-task unlearning approach. It employs a parameter self-sharing method to balance between forgetting the unlearned data and remembering the knowledge learned by the original model, where probability distributions are approximated by a neural network.

In addition to the aforementioned approaches, the work in \cite{wang2022federated} specializes in unlearning a specific type of partial data, particularly focusing on a category within the training dataset, i.e., a class. Typically initiated by the server, this unlearning process involves the application of quantization and pruning techniques. Specifically, the locally trained CNN model takes private images as input and produces feature map scores that assess the relationship between each channel and category. These scores are transmitted to the central server and aggregated into global feature map scores. The server utilizes TF-IDF to evaluate the relevance scores between channels and categories and creates a pruner to perform selective pruning on the most distinguishing channels of the target category. Subsequently, normal federated training proceeds with the exclusion of training data associated with the target category.

\textbf{Limitations.} 
Unlearning partial data through active unlearning is notably complex among all unlearning targets. It requires the removal of specific data while maintaining model performance on the remaining data. This complexity often results in more intricate algorithms, leading to increased computational and communication costs. Furthermore, this setting is susceptible to attacks based on over-unlearning, as identified in \cite{hu2023duty}, where adversaries can exploit the unlearning process to enhance the effects of remaining poisoned data, thus facilitating poisoning attacks. This vulnerability underscores the need for careful consideration in implementing active partial data unlearning.

\subsubsection{Unlearn entire client}

As mentioned earlier, unlearning through gradient ascent is a versatile approach suitable for both partial data and target client unlearning. As detailed in \cite{halimi2022federated}, the target client employs gradient ascent to maximize the local loss, subject to constraints determined by the reference model provided by the remaining clients. Specifically, during FL training, a client's objective is to address the following optimization problem:

$$\mathop{min}\limits_w \frac{1}{|D|}\sum\limits_{(x_i,y_i)\in D}\mathcal{L}(w;(x_i,y_i))$$

where $\mathcal{L}(w;(x_i,y_i))$ represents the loss function, which calculates the prediction error for an individual example $(x_i,y_i)$ from the dataset $D$, using the model parameters $w$. The unlearning method designed in \cite{halimi2022federated} is to reverse this learning process. That is, during unlearning, instead of learning model parameters that minimize the empirical loss, the client $i$ strives to learn the model parameters to maximize the loss. Additionally, to prevent the process of gradient ascent from producing an arbitrary model similar to a random model, the average of the other clients’ models, i.e., $w_{ref}=\frac{1}{N-1}\sum\limits_{i \neq j}w_j$, is used as a reference model, and an $\ell_2$-norm ball around the reference model is employed to limit the unbounded loss. Thus, during unlearning, the client solves the following optimization problem:
$$\mathop{max}\limits_{w:||w-w_{ref}||_2 \leq \delta} \frac{1}{|D_i|}\sum\limits_{(x_k,y_k)\in D_i}\mathcal{L}(w;(x_k,y_k))$$

Similarly to constraints, concurrent boost training on the remaining data \cite{wang2024server} and reference models \cite{bano2023federated} can also be adopted as mitigation methods. To reduce the computation overhead, unlearning is conducted based on gradient ascent on distilled dataset \cite{dhasade2023quickdrop}. Apart from the unlearning principle based on gradient ascent, 2F2L \cite{jin2023forgettable} and Appro-Fun \cite{xiong2024appro} adopt model scrubbing methods similar to previous works such as \cite{liu2022right} and \cite{cao2023fedrecover}. To enhance computational efficiency, the complex task of Hessian matrix inversions is approximated using a pre-trained deep neural network and Taloy expansion.

\textbf{Limitations.} Unlearning an entire client through active learning poses limitations in scenarios where the unlearning request comes from the client to be unlearned. In such cases, the client must remain in the system for unlearning, conflicting with the common ``request then leave" behavior in FL systems. Moreover, this method necessitates the unlearned client's continued participation alongside remaining clients, which is not cost-effective, especially in large-scale and resource-constrained FL systems. 

\begin{table}
\caption{A Summary of Active FU schemes}
\label{tab:fu_summary_table_continue_2}
\begin{centering}
\begin{tabular}{|>{\centering\arraybackslash}m{0.3cm}|>{\centering}m{0.5cm}|>{\centering\arraybackslash}m{1.3cm}|>{\centering}m{1.3cm}|>{\centering}m{1.3cm}|>{\centering}m{4cm}|>{\centering}m{1cm}|>{\centering}m{2.4cm}|}
\hline
 \cellcolor{mygray}  \textbf{\noun{}} &  \cellcolor{mygray} \textbf{\noun{Ref.}}&  \cellcolor{mygray} \textbf{\noun{Who-Unlearn}}&  \cellcolor{mygray} \textbf{\noun{Unlearn-What}}&  \cellcolor{mygray} \textbf{\noun{Principle}}  & 
 \cellcolor{mygray} \textbf{\noun{Method}} &
 \cellcolor{mygray} \textbf{\noun{Verifier}} &  \cellcolor{mygray}\textbf{\noun{Verify Method}} \tabularnewline
\hline
\hline

\parbox[t]{2mm}{\multirow{12}{*}{\rotatebox[origin=c]{90}{ \hspace{-10 cm} Active Unlearning}}}

&\cite{shaik2023framu}& Server \& All clients & Partial data & Fine-tuning & The server aggregates local models and attention scores from all clients, based on which filters out unlearned data points and updates the global model. & Server & Global model convergence
\tabularnewline \cline{2-8}

&\cite{zhu2023heterogeneous}& Server \& All clients & Partial data & Fine-tuning \& Multi-task unlearning & All clients iteratively optimize the local embedding based on mutual knowledge distillation following a multi-task style. & Server & Prediction results on knowledge
\tabularnewline \cline{2-8}

&\cite{che2023fast}& Server & Partial data & Model scrubbing & Treat unlearned data as a perturbation on the whole dataset. Refine the global model by random smoothing.& Server & Certified budget of data removals
\tabularnewline \cline{2-8}

&\cite{fan2022fast}& Server \& All clients & Partial data& Retraining & After each client removes unlearned data and calculates the new updates, the server identifies a prior state from which the remaining clients proceed with their training. & Server & Global model convergence
\tabularnewline \cline{2-8}




&\cite{xiongexact}& All clients& Partial data& Retraining& The client calculates a new model based on the remaining data. If the original model matches the new quantized model, the FL model remains unchanged. Otherwise, retraining is required. & NA &Accuracy-based metrics, MIA, speed-up ratio
\tabularnewline \cline{2-8}

&\makecell{\cite{gong2022forget} \\ \cite{gong2022compressed} \\ \cite{gong2023bayesian}}& Target client& Partial data& Retrain&  Use variational inference to approximate Bayesian posterior probability. After the client requests to leave, the client uses the remaining data to reapproximation the posterior probability and execute an extra round of model upload& NA&Accuracy, Posterior Distribution
\tabularnewline \cline{2-8}

&\cite{liu2020learn}& Target clients & Partial data & Model scrubbing & Dummy gradients are computed to align confidence vectors of the unlearned model with that of a perfectly unlearned model.  & NA & Forgetting rate
\tabularnewline \cline{2-8}

\hline
\hline

\end{tabular}
\par\end{centering}
\end{table}

\begin{table}
\caption{A Summary of Active FU schemes (continued)}
\label{tab:fu_summary_table_continue_3}
\begin{centering}
\begin{tabular}{|>{\centering\arraybackslash}m{0.3cm}|>{\centering}m{0.5cm}|>{\centering\arraybackslash}m{1.3cm}|>{\centering}m{1.3cm}|>{\centering}m{1.3cm}|>{\centering}m{4cm}|>{\centering}m{1cm}|>{\centering}m{2.4cm}|}
\hline
 \cellcolor{mygray}  \textbf{\noun{}} &  \cellcolor{mygray} \textbf{\noun{Ref.}}&  \cellcolor{mygray} \textbf{\noun{Who-Unlearn}}&  \cellcolor{mygray} \textbf{\noun{Unlearn-What}}&  \cellcolor{mygray} \textbf{\noun{Principle}}  & 
 \cellcolor{mygray} \textbf{\noun{Method}} &
 \cellcolor{mygray} \textbf{\noun{Verifier}} &  \cellcolor{mygray}\textbf{\noun{Verify Method}} \tabularnewline
\hline
\hline

\parbox[t]{2mm}{\multirow{12}{*}{\rotatebox[origin=c]{90}{ \hspace{-11 cm} Active Unlearning}}}

&\cite{wang2022federated}& Server and all clients & Partial data (a class)& Model scrubbing& The server assesses the relevance between channels and classes and establishes a pruner to selectively trim the most distinguishing channels of the target class. & NA& Accuracy-based, speed-up ratio, MIA
\tabularnewline \cline{2-8}

&\cite{xia2023fedme}& All clients & Partial data & Multi-task unlearning & Engage in multi-task learning to optimize the loss of the local model, the loss from an MIA-like evaluation model, and a penalty from the difference between the local model and the global model. & NA & Convergence analysis, accuracy-based metrics, forgetting rate
\tabularnewline \cline{2-8}

&\cite{ma2022learn}& Target client & Partial data & Model Scrubbing & Target clients iteratively minimizes the distance between the posteriors of the data to be forgotten and those of non-member data for unlearning. & NA & Accuracy-based and efficiency-based metrics, MIA
\tabularnewline \cline{2-8}

&\cite{li2023federated}& Target clients & Partial data & Multi-task unlearning & Synthetic labels are generated based on teacher ensembles for the data to be unlearned, and training is conducted using this data with synthetic labels to achieve unlearning. & NA & Accuracy, running time
\tabularnewline \cline{2-8}

&\cite{wang2023bfu}& Target client& Partial data& Multi-task unlearning& Adopts a multi-task unlearning approach that utilizes a parameter self-sharing method to strike a balance between forgetting the unlearned data and retaining the remaining knowledge.& NA & Running time, model differences, accuracy-based metrics, backdoor attack
\tabularnewline \cline{2-8}

&\cite{gong2021bayesian}& Target client& Partial data& Retraining & Shares a core concept with \cite{gong2022forget}, employing exponential family distributions in VI to approximate posterior probability.& NA&KL-divergence
\tabularnewline \cline{2-8}

&\cite{alam2023get}& Target clients & Partial data & Gradient ascent & Follows the gradient ascent method, utilizing the loss from the local benign dataset to constrain unbounded losses. & NA & Accuracy-based metrics, backdoor attack
\tabularnewline \cline{2-8}

&\cite{liu2022right}& Target client & Partial data & Model scrubbing & Scrubs the model based on the approximation of Hessian matrix using the remaining data. & NA & Running time, accuracy, model utility
\tabularnewline \cline{2-8}

\hline
\hline

\end{tabular}
\par\end{centering}
\end{table}

\begin{table}
\caption{A Summary of Active FU schemes (continued)}
\label{tab:fu_summary_table_continue_4}
\begin{centering}
\begin{tabular}{|>{\centering\arraybackslash}m{0.3cm}|>{\centering}m{0.5cm}|>{\centering\arraybackslash}m{1.3cm}|>{\centering}m{1.3cm}|>{\centering}m{1.3cm}|>{\centering}m{4cm}|>{\centering}m{1cm}|>{\centering}m{2.4cm}|}
\hline
 \cellcolor{mygray}  \textbf{\noun{}} &  \cellcolor{mygray} \textbf{\noun{Ref.}}&  \cellcolor{mygray} \textbf{\noun{Who-Unlearn}}&  \cellcolor{mygray} \textbf{\noun{Unlearn-What}}&  \cellcolor{mygray} \textbf{\noun{Principle}}  & 
 \cellcolor{mygray} \textbf{\noun{Method}} &
 \cellcolor{mygray} \textbf{\noun{Verifier}} &  \cellcolor{mygray}\textbf{\noun{Verify Method}} \tabularnewline
\hline
\hline

\parbox[t]{2mm}{\multirow{12}{*}{\rotatebox[origin=c]{90}{ \hspace{-11 cm} Active Unlearning}}}

&\cite{pan2022machine}& Target clients & Partial data & Retraining & Each client computes a new local vector, and these vectors are subsequently aggregated by the server. & Server &  Global model convergence
\tabularnewline \cline{2-8}

&\cite{jin2023forgettable}& Target client & Target client & Model scrubbing & Scrubs the model based on the approximation of Hessian matrix using public server data.& NA & Accuracy-based metrics, MIA
\tabularnewline \cline{2-8}

&\cite{halimi2022federated}& Target client & Target client & Gradient ascent & Target client computes the maximum empirical loss with the constraint of the reference model from remaining clients. & NA & Accuracy-based metrics, backdoor attack
\tabularnewline \cline{2-8}

&\cite{gu2024unlearning}& Server and all clients & Partial data & Model Scrubbing & Linear combination of the trained model and auxiliary model obtained during unlearning. & NA & Accuracy
\tabularnewline \cline{2-8}

&\cite{wang2024server}& Server & Target client & Gradient ascent & Unlearning low-quality data with concurrent boost training with good-quality data & Server & Accuracy, loss, running time
\tabularnewline \cline{2-8}

&\cite{zhang2023securecut}& Target client & Partial data & Retraining & Retrain the model based on prune local models& NA & Accuracy, loss
\tabularnewline \cline{2-8}

&\cite{dhasade2023quickdrop}& All clients & Target client or a class & Gradient ascent & Reverse training on the distilled dataset & NA & Accuracy, time, rounds, data size, MIA
\tabularnewline \cline{2-8}

&\cite{wang2024goldfish}& Server \& all clients & Partial data & Multi-task unlearning & Optimize model performance on the remaining dataset while considering bias caused by the unlearned data. & NA & Backdoor attack, L2 distance, JS divergence, T-test 
\tabularnewline \cline{2-8}

&\cite{zhao2023federated}& Server \& target clients & Partial data or a class & Fine-tuning & Fine-tuning based on randomly initialized degradation models & NA & Backdoor attack, accuracy
\tabularnewline \cline{2-8}

&\cite{wang2024efficient}& All clients & Partial data (a feature) & Retraining & Rapid retraining using first-order method based on reinitialized model.& NA & Accuracy-based metrics
\tabularnewline \cline{2-8}

&\cite{xiong2024appro} & Server \& target client & Target client & Model scrubbing & Achieving indistinguishability based on DP definition. & NA & Accuracy, loss, MIA, speed-up ratio
\tabularnewline \cline{2-8}

&\cite{xu2023revocation} & Server \& target client & Target client & Synthetic data  & Training on perturbed unlearned data. & NA & Accuracy
\tabularnewline \cline{2-8}

&\cite{gu2024ferrari} & Server \& target client & Target client & Model scrubbing  & Local feature unlearning to minimize feature sensitivity. & NA & Sensitivity, bias, backdoor attack
\tabularnewline \cline{2-8}

&\cite{meerzaconfuse} & Server \& target client & Target client or partial data & Multi-task unlearning  & Influence removal based on confusion Loss and performance recovery based on saliency map. & NA & Accuracy, MIA, backdoor attack
\tabularnewline \cline{2-8}

\hline
\hline

\end{tabular}
\par\end{centering}
\end{table}

\subsection{Verification}
\label{sec:verification}

In line with the description provided in VERIFI \cite{gao2022verifi}, the participant who requests unlearning, e.g., a target client or the server, is granted ``the right to verify" (RTV). This means that the requester has the ability to actively verify the unlearning effect after the unlearning process is completed. This section will provide an overview of the verification mechanisms proposed in existing FU works.

\subsubsection{Client-side}

To ensure the ``right to verify," it is imperative that federated unlearning schemes provide clients with the ability to confirm the successful unlearning of their data. Regrettably, this aspect has received limited attention in existing FU literature. In EMA \cite{huang2022dataset}, the target client employs various metrics, including correctness, confidence, and negative entropy, to assess the performance of the audited dataset concerning the global model. These metrics are then ensembled to determine whether they meet a predefined threshold, serving as an indicator of whether the target client's data has been effectively unlearned.

Another noteworthy contribution in this domain is VERIFI \cite{gao2022verifi}. VERIFI introduces two non-invasive verification methods that distinguish themselves from invasive techniques involving the injection of backdoors or watermarks, which manipulate the original data. In contrast, the verification methods proposed in VERIFI operate without modifying the data itself. These methods revolve around tracking a subset of unlearned data known as ``markers," selected based on specific criteria. The criteria encompass two primary categories: (i) Forgettable Memory, where markers are identified as a representative subset incurring a high variance of local training loss, and (ii) Erroneous Memory, which designates markers as incorrectly predicted samples labeled as erroneous. By actively monitoring the unlearned model's performance on these markers, clients can effectively verify whether the unlearning process has successfully removed the unlearned data.

\subsubsection{Server-side} The outcome of the verification conducted by the server plays an important role in determining when to stop the unlearning process within the FU system. For instance, FRAMU \cite{shaik2023framu} terminates unlearning when the difference between two consecutive global models becomes smaller than a predefined parameter. KNOT \cite{su2023asynchronous} concludes unlearning based on the required validation accuracy and the standard deviation across a recent history of such validation accuracies. ViFLa \cite{fan2022fast} and SCMA \cite{pan2022machine} end unlearning when the model converges. FedLU \cite{zhu2023heterogeneous} relies on prediction results derived from knowledge, while FFMU \cite{che2023fast} assesses whether data removals exceed a certified budget.

\subsubsection{Verification metrics}

While the remaining FU works reviewed in this survey do not explicitly introduce verification mechanisms, the verification metrics employed in these works to assess unlearning performance could provide valuable insights for future research. We have summarized these metrics adopted in reviewed FU works in Table \ref{tab:fu_summary_table}. 

Accuracy-based metrics over unlearned data are the most commonly utilized metrics in the reviewed FU works. For example, they are used in works such as \cite{wang2023bfu,jin2023forgettable,zhang2023fedrecovery,fraboni2022sequential,cao2023fedrecover,liu2021federaser,halimi2022federated,xia2023fedme,li2023federated,wu2023unlearning,alam2023get,liu2022right,zhu2023heterogeneous}. Metrics based on running time are employed to assess the efficiency of unlearning algorithms, as demonstrated in \cite{wang2023bfu,zhang2023fedrecovery,fraboni2022sequential,cao2023fedrecover,liu2021federaser,xia2023fedme,liu2020learn,ma2022learn,li2023federated,liu2022right,pan2022machine}. Furthermore, some works rely on verification through backdoor attacks, as in \cite{li2023subspace,cao2023fedrecover,yuan2023federated,halimi2022federated,ma2022learn,wu2023unlearning,alam2023get}, while others use membership inference attacks, as seen in \cite{jin2023forgettable,zhang2023fedrecovery,liu2021federaser,ma2022learn}. The difference between the unlearned model and the retrained model is adopted to evaluate unlearning performance in \cite{wang2023bfu,ma2022learn,gong2022compressed}.

\subsubsection{Limitations}

By surveying the existing FU literature, we can make several observations regarding verification as follows:

\begin{itemize}
\item Only a few research studies on FU take ``who-verify" into account, and the verification in almost all FU schemes remains at an experimental assessment level.
\item Most FU methods rely on the assumption that verification is conducted by the server rather than the clients.
\item There are no standard or widely adopted methods for proof of unlearning.
\end{itemize}

These observations indicate the need for research into "who-verify" and the development of efficient and robust verification methods conducted by different participants, especially client-verify methods. For instance, when considering MLaaS with unlearning services, clients must be allowed to verify if their data has been unlearned and its impact on the FL model has been removed. Only when the data removal adheres strictly to the specified unlearning request can the trustworthiness of the federated unlearning system be maintained.

\subsection{Lessons Learned}
In this section, we present the key lessons learned from our review of existing FU methods.

\begin{itemize}
    \item ``Who-unlearn" and ``Who-verify": According to the proposed taxonomy, unlearning can be carried out by the participant who initiates the unlearning request (e.g., client or server) or by other participants, excluding the one who made the request. Similarly, verification can also be performed by clients or the server. However, we observe that the alignment between ``Who-unlearn" and ``Who-verify" is not optimal in FU literature. In other words, an FU system should allow the participants who raise the unlearning request to either conduct the unlearning or perform the verification themselves. This ensures that the unlearning results are credible to the participants who made the request.
    \item Selection of unlearning principles: It can be observed that different unlearning principles vary in their reliance on access to the training data. For instance, gradient ascent-based unlearning methods heavily rely on the unlearned data, while fine-tuning-based methods may rely only on the remaining data, and retraining-based methods are more flexible. Therefore, selecting appropriate unlearning principles that align with the FU scenarios concerning data access levels should be carefully considered.
    \item Structure of unlearning requests: It is challenging to determine the structure of unlearning requests from the existing FU literature. The underlying assumption that the unlearner has direct access to the unlearned data appears contradictory to the privacy foundation of FL systems. For example, if the unlearner is the server and the unlearned data is held by the clients, this creates a conflict. The lack of consideration for the structure of unlearning requests and their integration within the FL system may hinder the adoption of unlearning services in a federated setting.
    \item Proof of unlearning: As mentioned earlier, there are no standard or widely adopted metrics for proof of unlearning. To deploy unlearning as a service within MLaaS, it is crucial to establish a standard, either globally or within regional organizations, to guide the design of unlearning verification. Additionally, emphasis should be placed on verification by the entity that raises the unlearning requests.
\end{itemize}

\section{FL-Tailored Optimizations, Limitations, and Applications}
\label{sec:fl_specific_considerations}

In addition to the primary challenges and solutions discussed in earlier sections, FU schemes face limitations due to the unique characteristics of the FL setting. These include (i) constrained resources, (ii) participant heterogeneity, and (iii) security and privacy threats. Furthermore, since FU systems involve additional unlearning processes compared to FL systems, new security and privacy threats arise. To address these issues, various optimization approaches and solutions have been proposed. In this section, we will delve into these limitations and concerns arising from the unique characteristics of the FL setting and explore the efforts made to address them. A summary of these FL-tailored optimization methods and solutions in FU is provided in Table \ref{tab:optimization-table}.

\begin{table*}[htbp!]
\centering
\caption{A Summary of FL-Tailored Optimization Methods in FU.}
\label{tab:optimization-table}
\begin{tabular}{|cc|c|c|}
\hline
\rowcolor[HTML]{9B9B9B} 
\multicolumn{2}{|c|}{\cellcolor[HTML]{9B9B9B}\textbf{Limitation}}                                                     & \textbf{Optimization Method}         & \textbf{Reference} \\ \hline \hline
\multicolumn{1}{|c|}{}                                              & \multirow{-1}{*}{Memory}                                 & Selective storage           & \cite{liu2021federaser,yuan2023federated,jiang2024towards}          \\ \cline{3-4} 
\multicolumn{1}{|c|}{}                                              &                                        & Compression          & \cite{lin2024scalable}         \\ \cline{2-4}
\multicolumn{1}{|c|}{}                                              &                                        & Size reduction              & \cite{xiongexact}              \\ \cline{3-4} 
\multicolumn{1}{|c|}{}                                              & \multirow{-1}{*}{Communication}        & Rounds reduction            &   \cite{xiongexact,fan2022fast,fraboni2022sequential}        \\ \cline{3-4}
\multicolumn{1}{|c|}{}                                              &                                        & Clustering   & \cite{su2023asynchronous,qiu2023fedcio,liu2024guaranteeing,wang2023fedcsa}         \\ \cline{2-4} 
\multicolumn{1}{|c|}{}                                              &                                        & Approximation               & \cite{jin2023forgettable,cao2023fedrecover,li2023federated,liu2022right}          \\ \cline{3-4} 
\multicolumn{1}{|c|}{}                                              &                                        & Parallel computation        &  \cite{che2023fast}         \\ \cline{3-4} 
\multicolumn{1}{|c|}{\multirow{-6}{*}{Constrained Resources}}       & \multirow{-1}{*}{Computation}          & Outsource computation       & \cite{li2023federated,cao2023fedrecover}          \\ \cline{3-4} 
\multicolumn{1}{|c|}{}                                              &                                        & \hspace{1em} Pre-computation        & \cite{gu2024unlearning}          \\ \cline{3-4}
\multicolumn{1}{|c|}{}                                              &                                        & \hspace{1em} Dataset distillation        & \cite{dhasade2023quickdrop}          \\ \hline
\multicolumn{1}{|c|}{}                                              & Partitioned feature           & Vertical FL                 &  \cite{liu2021revfrf, deng2023vertical,zhang2023securecut,wang2024efficient}        \\ \cline{2-4} 
\multicolumn{1}{|c|}{}                                              & Variational  representation         & Knowledge distillation      &  \cite{zhu2023heterogeneous,shaik2023framu,wang2024goldfish}         \\ \cline{2-4} 
\multicolumn{1}{|c|}{}                                              &                                        & Weighted aggregation        &   \cite{shaik2023framu, fan2022fast,wang2022federated,su2023asynchronous}        \\ \cline{3-4} 
\multicolumn{1}{|c|}{}                                              & \multirow{-2}{*}{Non-IID distribution} & Knowledge distillation      &  \cite{zhu2023heterogeneous, ye2023heterogeneous}          \\ \cline{2-4} 
\multicolumn{1}{|c|}{\multirow{-4}{*}{Participant Heterogeneity}}   & Training capability                    & Client clustering           &  \cite{su2023asynchronous}         \\ \cline{2-4}
\multicolumn{1}{|c|}{}                                              &  Diverse payoff                                      & \hspace{1em} Incentive mechanism      & \cite{ding2023incentive,lin2024incentive,ding2023strategic}          \\ \hline
\multicolumn{1}{|c|}{}                                              &                                        & Indirect information        & \cite{pan2022machine,ding2023incentive,ye2023heterogeneous}         \\ \cline{3-4} 
\multicolumn{1}{|c|}{\multirow{-2}{*}{Security \& Privacy Threats}} & \multirow{-2}{*}{Privacy-preservation} & PETs &  \cite{liu2024privacy,liu2021revfrf,liu2020learn,xiongexact,zhang2023fedrecovery,liu2024guaranteeing,zhang2023securecut}           \\ \hline \hline
\end{tabular}
\end{table*}

\subsection{Constrained Resources} In cross-device FL settings, FL clients are typically resource-constrained mobile devices that may drop out of the system at any time \cite{kairouz2021advances}. As shown in the FU workflow in Figure \ref{fig:fu-workflow}, federated unlearning is a subsequent process after federated learning. Therefore, the characteristic of resource-constrained participants exists in both FL and FU systems. Consequently, it is crucial to consider the resource requirements and consumption of FU schemes.

\subsubsection{Memory} In many FU works, historical information is essential for facilitating the unlearning process. Nevertheless, this implies that the server needs to store a substantial volume of data, leading to significant memory consumption. This historical information can be gradients and global models \cite{cao2023fedrecover,fraboni2022sequential,liu2021federaser,yuan2023federated,wu2023unlearning,halimi2022federated}, gradient residuals \cite{zhang2023fedrecovery}, specific state \cite{shaik2023framu}, or some intermediate results \cite{liu2021revfrf}.

\textbf{Approaches to mitigation.} Memory consumption reduction can be achieved by selectively storing historical information. For instance, in FedEraser \cite{liu2021federaser}, the server stores clients' gradients at specific intervals of FL rounds or based on the importance of gradients \cite{jiang2024towards}. Similarly, in FRU \cite{yuan2023federated}, only important updates to clients' item embeddings are stored. In addition, by adopting coding-based techniques, the storage can be further compressed as demonstrated in \cite{lin2024scalable}.

\subsubsection{Communication} Unlearning in the FU system typically necessitates FL clients to transmit extra information, such as gradients, to the server to facilitate the process. For example, in FRAMU \cite{shaik2023framu}, clients additionally send attention scores to the server. In FedLU \cite{zhu2023heterogeneous}, loss information is transferred for mutual knowledge distillation between the server and clients. In SFU \cite{li2023subspace}, the remaining clients need to send their representation matrix to the server. In \cite{halimi2022federated}, the remaining clients' models are sent to the target client as references.

\textbf{Approaches to mitigation.} Reducing the communication cost can be achieved by minimizing the size of the model to be transferred, which may involve methods like quantization \cite{xiongexact}, and by reducing the number of unlearning rounds. For retraining-based unlearning, the FU system may roll back the global model to a state where it has not been significantly influenced by the target client. From this point, all FL clients can conduct the retraining process \cite{xiongexact,fan2022fast,fraboni2022sequential}, thus reducing the unlearning rounds and enhancing communication efficiency. Besides, clustering is used to divide FL users into groups, each with its own model. The final inference is determined by a majority vote from these sub-models. This method confines unlearning processes to individual clusters, eliminating the need for participation from all users, thus improving efficiency \cite{su2023asynchronous,qiu2023fedcio,liu2024guaranteeing,wang2023fedcsa}.

\subsubsection{Computation} FU often requires clients to engage in additional computational tasks compared to standard FL. These tasks can involve generating dummy gradients \cite{liu2020learn}, conducting online reinforcement learning \cite{shaik2023framu}, seed model generation \cite{ye2023heterogeneous}, or computing Hessian matrices \cite{jin2023forgettable,cao2023fedrecover,li2023federated,liu2022right}. It's important to note that some of these computational tasks can be resource-intensive, which could pose challenges for deploying unlearning mechanisms in FL systems.

\textbf{Approaches to mitigation.} To enhance computational efficiency, approximation methods are commonly employed to accelerate certain components of FU algorithms. For example, the computation on the Hessian matrix can be approximated using techniques like a pre-trained deep neural network with Taylor expansion \cite{jin2023forgettable}, the L-BFGS algorithm \cite{cao2023fedrecover}, or the Fisher information matrix \cite{li2023federated,liu2022right}. Additionally, other optimizations are based on different FU structures. For instance, some works conduct training and unlearning simultaneously \cite{che2023fast}, similar to the approach used in \cite{zhang2022prompt} in an MU setting. Orthogonally, computation tasks can be outsourced to a trusted third party, as demonstrated in \cite{li2023federated}. Another approach is transferring the majority of tasks to the server for estimation while executing only a small part of computation tasks for calibration \cite{cao2023fedrecover}, or reducing the computational tasks in the unlearning process by involving some pre-computation during the learning phase \cite{gu2024unlearning}. In \cite{dhasade2023quickdrop}, dataset distillation is adopted to compress the size of the dataset while preserving the unlearning performance, hence reducing the computational overhead.

\subsection{Participant Heterogeneity} In both FL and FU systems, clients exhibit heterogeneity in various aspects, encompassing differences in data structures, data distributions, such as vertical partitioned features \cite{liu2021revfrf, deng2023vertical}, variational data representations \cite{shaik2023framu, zhu2023heterogeneous}, and the presence of Non-IID data \cite{shaik2023framu, fan2022fast, wang2022federated, su2023asynchronous}. Furthermore, there are disparities in training capabilities on computational, communication and memory, with some clients operating on resource-constrained mobile or IoT devices \cite{su2023asynchronous}. The existence of such diversity highlights the importance of developing heterogeneity-aware approaches for federated unlearning.

\textbf{Approaches to mitigation.} To address challenges associated with vertical partitioned features, certain vertical federated learning schemes are employed, as seen in works like \cite{liu2021revfrf,deng2023vertical,zhang2023securecut,wang2024efficient}. To handle Non-IID data distributions and variations in data representations, weighted aggregation techniques are commonly utilized, leveraging different metrics such as attention-based mechanisms \cite{shaik2023framu, fan2022fast}, TF-IDF \cite{wang2022federated}, and model sparsity \cite{su2023asynchronous}. The introduction of knowledge distillation techniques helps mitigate issues arising from data heterogeneity, bias, and diverse model architectures \cite{zhu2023heterogeneous,wang2024goldfish,ye2023heterogeneous}. Additionally, clustering based on local computational resources is considered to achieve asynchronous aggregation for FL, along with clustered retraining for FU \cite{su2023asynchronous}. Furthermore, incentive mechanisms can be adopted in FU systems to deal with diverse payoffs for different FU participants \cite{ding2023incentive,lin2024incentive,ding2023strategic}.

\subsection{Security and Privacy Threats}

Privacy and security issues in FU systems encompass those present in FL, such as the risk of information leakage and both targeted and untargeted attacks on ML models (see Section \ref{sec:preliminaries_and_backgrounds} for more details). For instance, the leakage from gradients even allows the attacker to recover images with pixel-wise accuracy and texts with token-wise matching \cite{zhu2019deep}. Methods to mitigate such a risk focus on privacy-preserving aggregation \cite{zheng2022aggregation,liu2022efficient,liu2022privacy}. Additionally, recent research highlights that malicious clients can launch (i) untargeted poisoning attacks, which aim to slow the learning process or reduce the global model's performance \cite{shejwalkar2022back,liu2020poisoning}, or (ii) targeted backdoor attacks, where a backdoor is embedded into the model, triggering malicious behavior under specific input conditions \cite{jebreel2023fl,li2021hidden,huang2023training}. Such attacks can quickly degrade the global model's performance or implant backdoors within a few FL rounds, with effects lasting for many rounds, posing serious security risks \cite{ma2023quantization,zhang2023fltracer,li2023ntd}.

\textbf{Approaches to mitigation.} To address these security and privacy concerns, mitigation strategies involve the use of indirect information, such as transmitting representative vectors instead of centroids in federated clustering \cite{pan2022machine}, calculating clients' contributions using federated Shapley values \cite{ding2023incentive}, and generating predictions on the ensemble of seed models acquired through knowledge distillation \cite{ye2023heterogeneous}. Moreover, integrating privacy-enhancing techniques (PETs) into the FL-FU workflow can bolster security and privacy guarantees, such as employing secure random forest construction for secure random forest re-construction \cite{liu2021revfrf}, secure aggregation \cite{liu2024dynamic} for privacy-preserving gradient sum-up \cite{liu2024guaranteeing,liu2020learn}, secure two-party computation for privacy-preserving unlearning \cite{liu2024privacy}, homomorphic encryption for initialization \cite{zhang2023securecut}, and differential privacy mechanisms \cite{zhang2023fedrecovery} for rendering unlearned mode indistinguishable from the retrained one.

\subsection{Applications for Enhanced Security}

In addition to ensuring RTBF, federated unlearning has significant applications in enhancing the security and integrity of federated learning models. In the context of poisoning recovery, it enables the removal of maliciously inserted data from trained models, thus restoring their original accuracy and reliability \cite{cao2023fedrecover}. For backdoor removal, federated unlearning is instrumental in eliminating hidden backdoors in FL models \cite{alam2023get,wu2023unlearning,wu2023unlearning}, which could otherwise be exploited for adversarial purposes. Additionally, it plays a crucial role in addressing data misuse in unauthorized training by enabling the removal of improperly used data or outdated data \cite{fan2022fast,ye2023heterogeneous}, thereby ensuring compliance with standards and regulations. These applications underscore federated unlearning's importance in maintaining the trustworthiness and security of machine learning models.

\subsection{Lessons Learned}
In this section, we present the key lessons learned from our review of FL-tailored optimizations in existing FU methods.

\begin{itemize}
    \item Trade-offs of resource consumption: The interplay between memory, communication, and computation is complex. We observe independent efforts to optimize efficiency in each of these areas within existing FU approaches. However, there is a lack of combined consideration, which is crucial, especially for resource-constrained FL participants. The trade-offs between memory, communication, and computation should be thoroughly investigated to achieve optimal results.
    \item Consideration of participant heterogeneity: We observe that a few studies consider the heterogeneity among FL participants, but this area still requires further exploration. For instance, when managing heterogeneity based on training ability, memory, communication, and computation should all be taken into account. Additionally, existing FU literature primarily addresses simple Non-IID settings with basic data representations. There is a need to investigate complex Non-IID data with other representations, such as graphs.
    \item Security \& privacy: More studies on machine unlearning reveal that additional privacy leakage can occur in the unlearning setting compared to the learning process. In addition, malicious unlearning, where attackers raise crafted unlearning requests to achieve goals such as degrading model performance or injecting backdoors, exists. However, there is a lack of extended investigation into these issues and the development of defense strategies in the federated setting.
\end{itemize}

\section{Discussions and Promising Directions}
\label{sec:discussions_and_promising_directions}

In previous sections, we conducted a comprehensive survey of FU schemes. However, given the rapid evolution of FU schemes and their increasing deployment, numerous emerging challenges and open problems are awaiting further investigation. Many of these challenges necessitate additional properties and broader capabilities from FU schemes. In this section, we extend our discussion to encompass these challenges and present potential research directions, highlighting areas where FU schemes can further enhance their capabilities.

\subsection{Privacy-preserving FU}
The majority of FU schemes reviewed in this survey heavily rely on gradient information from the target client or all clients. For instance, historical client models and updated client models are exposed to the server in various schemes \cite{zhang2023fedrecovery,liu2021federaser,yuan2023federated,cao2023fedrecover}. However, it has been highlighted that with only a client's model and the global model, an attacker, such as a malicious central server, can accurately reconstruct a client's data in a pixel-wise manner for images or token-wise matching for texts, as discussed in \cite{zhu2019deep}. To counteract this ``deep leakage from gradients," privacy-preserving techniques (PPT), such as Homomorphic Encryption (HE) \cite{gentry2009fully}, Multi-Party Computation (MPC) \cite{yao1982protocols}, and Differential Privacy (DP) \cite{dwork2014algorithmic}, can be integrated to aggregate clients' locally trained models in a privacy-preserving manner. However, it's important to note that this approach significantly impacts the performance of existing FU algorithms, as the server no longer has access to the gradient of the target client. Therefore, there is a critical need for privacy-preserving FU methods that enable unlearning while preserving clients' data privacy. Additionally, as previously mentioned, there are additional security and privacy risks introduced in MU systems due to information leakage from the differences between the original and unlearned models \cite{chen2021machine}. This potential information leakage must also be analyzed within a FU system, and corresponding defense mechanisms should be developed.

\subsection{Verification and proof of unlearning}

As elaborated in Section \ref{sec:verification} and summarized in Table \ref{tab:fu_summary_table}, it is unfortunate that this aspect regarding ``who-verify" has received limited attention in existing FU literature. Given that in most real-life FL systems, unlearning requests are typically initiated by a specific target FL client, there should be a heightened emphasis on client-side verification that allows clients to verify if their data has been unlearned and its impact on the FL model has been removed. This approach not only enhances privacy guarantees \cite{wang2023federated} but also aligns with the marking-then-verification strategy outlined in \cite{gao2022verifi}, hence maintaining the trustworthiness of the federated unlearning system. Traditional invasive marking methods, including watermarking, fingerprinting, and backdoor attacks, manipulate the original data, potentially impacting the performance of the FL model. Consequently, the exploration of effective and non-invasive verification mechanisms is a critical area of research within the context of FU.

Apart from verification mechanisms, the development of ``proof of unlearning" using applied cryptography such as Zero-Knowledge Proofs (ZKPs) or Trusted Execution Environments (TEEs) presents a compelling research area, particularly for environments where mentioned unlearning verification methods are impractical or trust is limited. This approach offers enhanced cryptographic security guarantees, ensuring more robust and verifiable federated unlearning in sensitive or distrustful settings. Emphasizing cryptographic and hardware-based solutions marks a critical step forward in secure and trustworthy ML practices.

\subsection{Emerging threats in unlearning}
Due to the nature of unlearning, additional security and privacy risks are introduced in FU systems. For instance, the information can be leaked by the differences between the original and unlearned models \cite{chen2021machine, hu2024learn, gao2022deletion, lu2022label}. This could exacerbate client privacy issues if an attacker has access to the model before and after the unlearning. Furthermore, from a security perspective, various studies demonstrate that adversarial users can submit crafted unlearning requests with untargeted goals, such as degrading the utility of the unlearned model \cite{hu2024duty,zhao2024static,qian2023towards}, or untargeted goals, such as injecting backdoors \cite{liu2024backdoor,di2022hidden,zhao2024static,qian2023towards}. These issues highlight potential vulnerabilities in FU schemes. Research needs to focus on developing robust defense mechanisms to mitigate these risks and ensure the integrity and security of FU systems.

\subsection{Awareness of client-dynamics}
The process of federated unlearning introduces significantly more non-determinism compared to centralized machine unlearning. This increased complexity arises from the random selection of clients and data for global aggregation and local training in each round, as well as the presence of potentially numerous dropped and newly joined clients. The unlearning process becomes even more challenging when considering the need to recall past clients for unlearning and retraining. This is particularly difficult for more complex FU schemes, such as privacy-preserving FU, where the integration of PETs must also provide resilience to dynamic client participation. Addressing these dynamic challenges requires the development of client-dynamics-aware FU algorithms. Such algorithms must be capable of adapting to the fluid nature of client involvement in FU, ensuring the integrity and effectiveness of the unlearning process even as clients frequently join or leave the network. This area of research is crucial for the advancement of FU, aiming to create robust solutions that maintain high standards of privacy and efficiency despite the inherent non-determinism of federated environments.

\subsection{Domain-specific applications}
Machine unlearning techniques are employed in diverse scenarios, such as LLMs \cite{pawelczyk2023context,liu2024towards,si2023knowledge,kumar2022privacy}, recommendation systems \cite{li2024making,chen2022recommendation,li2023selective}, and specific application scenarios such as health \cite{fraboni2023validation,elbedoui2023ecg,zhong2024federated}, IoT and blockchain \cite{yuan2024towards,zuo2024federated,wang2023edge,liu2024decentralized,lin2024blockchain}, HAR \cite{chen2024federated}, and metaverse and digital twin \cite{wang2023mitigating,islam2024federated}, to adhere to data privacy and compliance objectives for RTBF. In LLMs, unlearning helps ensure that sensitive or outdated information can be effectively removed, maintaining user privacy and data accuracy. For recommendation systems, unlearning allows for the deletion of user-specific data upon request, thereby enhancing user trust and compliance with privacy regulations. These techniques are also utilized in graph neural networks (GNNs) \cite{wu2023graphguard,chien2022certified,chen2022graph}, addressing privacy concerns in GNNs \cite{wu2023securing,zhang2023demystifying,wu2022model} to ensure data accuracy and relevance, and knowledge graphs to erase specific knowledge \cite{liu2024federated}. These applications demonstrate the versatility and critical role of machine unlearning in various technological domains. However, these domain-specific MU applications have yet to be extensively adapted to FU in federated settings. This gap underscores the potential for expanding the scope and applicability of FU strategies to these areas, encouraging further research into adapting these unlearning techniques for federated learning environments.

\subsection{Fairness and explainability}
Researching fairness and explainability in FU algorithms is essential, given the intricacies of ML and the distributed nature of FL. For instance, overlapping data among different FL clients is a common scenario. Unlearning data in such overlaps might fulfill the unlearning request from one client but could adversely affect the performance of other clients sharing that data. Furthermore, unlearning in FL introduces an extra layer of complexity, complicating the understanding of the model's alterations and their impact on the system as a whole. Addressing these issues is crucial to ensure transparency, trustworthiness, and adherence to regulations. Bridging the knowledge gap in this research field is imperative for improved decision-making interpretation in distributed AI systems. This is especially vital in industries where the processes of learning and unlearning carry profound ethical and legal connotations. Additionally, unlearning can be utilized as a method to enhance fairness in federated learning (FL), which is often a challenge due to the Non-IID nature of the data across different clients. By selectively unlearning biased or unfair data contributions, the overall model can be adjusted to provide more equitable outcomes, addressing the inherent discrepancies that arise from Non-IID data distributions. Research in this field is promising as unlearning not only improves the fairness of the FL model but also ensures that the model's performance is more consistent and reliable across diverse data sources.

\subsection{Integration with MLaaS}
A key future direction for FU lies in its integration with MLaaS. This involves addressing several critical challenges. First, the structure of unlearning requests needs to be carefully designed to meet the stringent privacy requirements of FL. Effective protocols must ensure that unlearning processes do not compromise client privacy. Second, the system must efficiently handle multiple unlearning requests. This includes developing strategies to manage these requests in a scalable and responsive manner. Third, maintaining the quality of service (QoS) in MLaaS is essential, particularly regarding throughput and privacy guarantees. One challenge is that executing unlearning requests might necessitate halting inference services, impacting QoS. Conversely, not performing unlearning contravenes the RTBF regulation. Therefore, future research should focus on creating mechanisms that respect RTBF while minimizing service disruptions. This would enable MLaaS to deliver reliable, privacy-conscious, and high-performance machine learning services.

\section{Conclusions}
\label{sec:conclusions}

In conclusion, this survey has made remarkable contributions to the field of federated unlearning. We began by meticulously formalizing the targets and challenges of federated unlearning and introducing an innovative unified federated unlearning workflow. We then derived a novel taxonomy for existing federated unlearning methods, based on crucial factors such as who initiates the unlearning, what precisely needs to be unlearned, and how to effectively verify the unlearning results in federated settings. Furthermore, we thoroughly explored various optimizations tailored to federated learning and provided a critical examination of their limitations. Through these comprehensive efforts, we have gained profound insights into the current challenges in federated unlearning and have outlined promising research directions for the future. This survey stands as an invaluable and insightful resource for researchers and practitioners, significantly advancing the rapidly evolving field of federated unlearning.

\begin{acks}
This research is supported by the National Research Foundation, Singapore and Infocomm Media Development Authority under its Trust Tech Funding Initiative. Any opinions, findings and conclusions or recommendations expressed in this material are those of the author(s) and do not reflect the views of National Research Foundation, Singapore and Infocomm Media Development Authority.
\end{acks}

\bibliographystyle{ACM-Reference-Format}
\bibliography{sample-base}


\end{document}